\documentclass[journal,draftcls,onecolumn,12pt,twoside]{IEEEtran}
\usepackage{amsfonts, amssymb}
\usepackage{amsmath}
\usepackage{cite}
\usepackage[pdftex]{graphicx}
\usepackage{algorithmic}
\usepackage[T1]{fontenc}
\usepackage{balance}%
\usepackage{amsfonts}%
\usepackage{amssymb}
\usepackage{xcolor}
\usepackage{lipsum}
\usepackage{mathtools}
\usepackage{cuted}
\usepackage{amsthm}
\usepackage{subcaption}
\usepackage{float}
\usepackage[font=small, labelsep=none]{caption}
\usepackage{multicol}
\usepackage{graphicx}
\usepackage{flushend}

\usepackage{caption}
\usepackage{booktabs}
\usepackage{pifont}
\usepackage{tabularx,colortbl}
\usepackage{colortbl}
\usepackage{ctable}
\usepackage{algorithm,algorithmic}

\providecommand{\U}[1]{\protect\rule{.1in}{.1in}}
\begin{document}
\title{\LARGE {Zero-Energy} RIS-Assisted Communications With Noise Modulation and Interference-Based Energy Harvesting}

\author{Ahmad~Massud~Tota~Khel,~Aissa~Ikhlef,~\IEEEmembership{Senior~Member,~IEEE}, ~Zhiguo~Ding,~\IEEEmembership{Fellow,~IEEE},~and~Hongjian~Sun,~\IEEEmembership{Senior~Member,~IEEE}\thanks{This work was supported by the CHEDDAR: Communications Hub for Empowering Distributed ClouD Computing Applications and Research funded by the UK EPSRC under grant numbers EP/Y037421/1 and EP/X040518/1.}\thanks{Manuscript received xxxxx xx, 2025; revised xxxxxx xx, 2025;
accepted xxxxxxx xx, 2025. Date of publication xxxxxx xx, 2025; date
of current version xxxxxxx xx, 2025. The associate editor coordinating the
review of this manuscript and approving it for publication was xxxxxxx xxxxxxxx.
\it{(Corresponding author: Ahmad Massud Tota Khel.)}}
\thanks{Ahmad Massud Tota Khel, Aissa Ikhlef, and Hongjian Sun are with the Department of Engineering, Durham University, DH1 3LE Durham, U.K. (e-mail:
ahmad.m.tota-khel@durham.ac.uk; aissa.ikhlef@durham.ac.uk; hongjian.sun@durham.ac.uk).}\thanks{Zhiguo Ding is with the Department of Electrical and Electronic Engineering, University of Manchester, M13 9PL Manchester, U.K. (e-mail: zhiguo.ding@manchester.ac.uk). }%
}
\maketitle

\vspace{-30pt}\begin{abstract}
To advance towards carbon-neutrality and improve the limited {performance} of conventional passive wireless communications, in this paper, we investigate the integration of noise modulation with zero-energy reconfigurable intelligent surfaces (RISs). In particular, the RIS reconfigurable elements (REs) are divided into two groups: one for beamforming the desired signals in reflection mode and another for harvesting energy from interference signals in an absorption mode, providing the power required for RIS operation. Since the harvested energy is a random variable, a random number of REs can beamform the signals, while the remainder blindly reflects them. We present a closed-form solution and a search algorithm for REs allocation, jointly optimizing both the energy harvesting (EH) and communication performance. Considering the repetition coding technique and discrete phase shifts, we derive analytical expressions for the energy constrained success rate, bit error rate, optimal threshold, mutual information, {and energy efficiency}. Numerical and simulation results confirm the effectiveness of the algorithm and expressions, demonstrating the superiority of the proposed integration over conventional noise-modulation systems. It is shown that by properly allocating the REs, both the EH and communication performance can be improved in low to moderate interference scenarios, while the latter is restricted in the high-interference regime.
\end{abstract} 

\markboth{IEEE Transactions on Green Communications and Networking}{Tota Khel \MakeLowercase{\textit{et al.}}: Zero-Energy RIS-Assisted Communications With Noise Modulation ...}
\vspace{-10pt}
\IEEEpeerreviewmaketitle

\section{Introduction}
\vspace{-10pt}To advance towards carbon-neutrality and contribute to meeting the global warming and carbon emission targets set by the Paris Agreement and International Telecommunication Union (ITU)\cite{worldbank}, energy-efficient technologies must be deployed in future wireless systems. Passive wireless communications, e.g., backscatter and noise modulation, have been proven to be energy-efficient solutions for {Internet-of-Things} (IoT) communications \cite{9518381,9034159,10269022,kish2005stealth,kapetanovic2022communication,kapetanovic2023cosmic,10373568,9980386,10706852,10475689,kish2006totally}. {In backscatter communications, information bits are conveyed using radio frequency (RF) signals by modulating the reflection characteristics of passive transmitters (Txs) \cite{9518381,9034159,10269022}. In contrast, thermal noise modulation-based communications encode information by modulating the naturally occurring thermal noise generated by passive electronic components—such as resistors and capacitors—without relying on external power sources or RF signals \cite{kish2005stealth,kapetanovic2022communication,kapetanovic2023cosmic,10373568,9980386,10706852}. Specifically, the Tx switches between different states (e.g., low-value and high-value resistors), typically using a digital logic controller or a simple switching circuit, to modulate the variance (or power spectral density) of the emitted thermal noise. This change in variance represents binary information—such as bit-0 and bit-1—enabling data transmission through passive statistical modulation of noise. Since thermal noise power is directly proportional to resistance, according to the Johnson–Nyquist noise principle, the receiver can distinguish between transmitted bits by analyzing the statistical properties (e.g., variance or energy) of the received signal\cite{10373568,9980386}.} 

Although both {backscatter and noise modulation systems} use passive Txs, backscatter systems suffer from double path-loss and fading, i.e., RF source$\to$Tx$\to$receiver (Rx) \cite{9518381,9034159,10269022}. Moreover, due to resistance to eavesdropping and hardware simplicity, noise modulation-based systems have been envisioned as a promising and cost-effective solution for future IoT communications\cite{kish2005stealth,9980386,kapetanovic2022communication,10475689,kish2006totally}. {While thermal noise modulation enables zero-energy transmission, its implementation, similar to other passive wireless systems, requires careful consideration of hardware-induced limitations. For instance, low-power signals generated by an unbiased resistor/capacitor switching circuit make the system highly susceptible to fading and additive white Gaussian noise (AWGN), requiring ultra-sensitive transceivers for reliable detection—an implementation that is difficult to realize under cost and complexity constraints \cite{9980386}. These hardware limitations significantly restrict the achievable data rate, Tx-Rx communication range, and overall robustness \cite{9980386}. Furthermore, the lack of conventional deterministic signals in noise modulation, combined with the hardware-induced constraints, makes it difficult to acquire accurate channel state information (CSI) \cite{10742895}.}

{To address the performance limitations of noise modulation systems, integrating a reconfigurable intelligent surface (RIS)—an energy-efficient technology composed of numerous reconfigurable elements (REs)—offers a promising solution by enabling a reconfigurable propagation environment \cite{10348506,9866567,10356096,10328394,9440664,10438079}.} By intelligently tuning the properties of the REs, such as phase shifts, the incident signals can be focused towards a desired destination, improving the receive signal strength, communication range, and overall performance \cite{10348506,9866567,10356096,10328394}. Although the powers required for the phase shift configurations and the RIS controller are remarkably lower than that of conventional relays, a dedicated power source is still required, limiting the benefits of RISs\cite{9866567,10356096}. Alternatively, a {zero-energy} RIS can be achieved by implementing energy harvesting (EH) techniques. This involves allocating some REs to operate in {the} absorption mode, allowing them to harvest energy from existing RF signals to power the REs in {the} reflection mode for beamforming, the RIS controller, and the EH circuitry\cite{10348506,9866567,10356096}. Thus, by integrating {zero-energy} RISs with noise modulation-based systems, energy-efficient IoT communications with improved reliability and extended communication range can be achieved. 

{It is noteworthy that effective beamforming requires CSI at the RIS, and the existing literature reports various estimation methods that can be applied to RIS-assisted noise modulation systems. For instance, in \cite{9518381}, the CSI was acquired for RIS-assisted backscatter systems in short-range and stationary environments via RIS training and phase control. Thus, given that both backscatter and noise modulation systems use similar passive Txs and low-cost Rxs, along with low-complexity signal processing \cite{10475689}, this method can be applied to estimate the CSI. In \cite{10742895}, a pilot-assisted CSI estimation method for noise modulation systems was proposed, which can be adapted for the considered integration. Considering the short-range nature of noise modulation and assuming a stationary environment, the position-based CSI acquisition method used in \cite{10591780} is also applicable to estimate the CSI. Furthermore, the method proposed in \cite{9726785}—where a RIS equipped with an RF chain estimates the CSI by receiving pilot signals and applying structured phase shifts to determine the incident signal’s direction of arrival—can be applied to the proposed system. } 
\vspace{-23pt}\subsection{Related Works}\vspace{-7pt}
Several studies have examined the transceiver hardware design and performance of noise modulation systems without integrating RISs\cite{kish2005stealth,kapetanovic2022communication,kapetanovic2023cosmic,10373568,9980386,10706852,10475689,kish2006totally}. The authors in \cite{kish2005stealth} proposed a novel concept of noise modulation that utilizes the noise power to send information securely. They demonstrated that by using thermal noise for classical communications and vacuum fluctuations for quantum communications, it is possible to achieve secure wireless systems. The authors in \cite{kapetanovic2022communication} developed the hardware design and experimentally assessed the throughput and communication range of conventional noise modulation systems, without considering the fading effects, in which information bits were modulated by thermal noise produced by a resistor and an open or short circuit. In \cite{kapetanovic2023cosmic}, the hardware design of noise modulation-based cosmic backscatter systems was demonstrated, where information bits were modulated using resistors under {conditions with various temperatures}. The bit error rate (BER) performance of conventional noise modulation systems was studied in \cite{10373568,9980386,10706852}, where the analytical BER expressions in \cite{10373568} and \cite{9980386} are based on the classical Gaussian approximation, {and the effects of fading have been ignored in} \cite{9980386} and \cite{10706852}. In \cite{10475689}, the transceiver hardware design of a backscatter system was demonstrated, where information bits were modulated and transmitted using the noise modulation technique without the need for pre-existing RF signals. The study also confirmed the hardware simplicity and robustness of the noise modulation-based system compared to conventional backscatter systems. In addition, a secure noise modulation-based system design was proposed in \cite{kish2006totally}, utilizing the Kirchhoff-law-Johnson-noise method with two pairs of resistors at the {transceivers}.

On the other hand, while RISs with dedicated energy sources have been extensively studied in the literature, few studies have examined {zero-energy} RIS-assisted systems without utilizing the noise modulation technique\cite{10348506,9866567,10356096}. For instance, in \cite{10348506} and \cite{9866567}, {zero-energy} RIS-assisted conventional systems under different configurations were considered, where a group of REs operating in {the} absorption mode harvests energy from a dedicated {signal source}, rather than from pre-existing ambient RF or interference signals. Assuming a fixed number of REs for beamforming and ignoring phase shift errors and interference, the authors in \cite{10348506} analyzed the joint energy-data rate outage probability and energy efficiency {(EE)}, while the authors in \cite{9866567} proposed algorithms for REs allocation to maximize the signal-to-noise ratio (SNR). The authors of \cite{10356096} proposed sequential phase-alignment algorithms to maximize the harvested energy and achieve a {zero-energy} RIS with a fixed number of REs; however, the communication performance remains unexplored. {In particular, the approach proposed in \cite{10356096}} aims to either harvest sufficient energy for the {RIS} from a single ambient RF source or reduce its dependence on dedicated energy sources when the harvested energy is insufficient.
\vspace{-15pt}\subsection{Motivations and Contributions}\vspace{-5pt}
 {As discussed earlier, numerous studies have investigated noise modulation and RISs individually. Prior research has shown that noise modulation enables zero-energy transmission; however, the resulting low-power signals and the randomness of wireless environments severely restrict the system reliability, communication range, and overall performance. Meanwhile, those existing on RISs have demonstrated that the signal quality, communication range, and overall performance can be significantly enhanced without increasing the Tx power. These insights suggest that RISs could mitigate the inherent performance limitations of noise modulation, positioning their integration as a promising and energy-efficient solution for IoT communications—an area that remains unexplored. In addition, existing works on noise modulation primarily focus on hardware implementation or basic BER analysis under idealized conditions, while zero-energy RIS-assisted designs typically assume dedicated energy sources and ideal phase shifts.}

{Motivated by these factors,} this paper investigates the integration of noise modulation and zero-energy RISs as an energy-efficient solution for IoT communications. {Unlike prior works,} we explore the use of co-existing interference signals—typically considered harmful in propagation environments—as potentially useful energy sources for EH at the RIS, eliminating the need for dedicated EH signals. Moreover, previous works on zero-energy RISs assume a fixed number of REs with ideal beamforming and phase shifts, but in practice: (i) REs have discrete phase shifters, leading to non-ideal beamforming and phase quantization errors, and (ii) when the harvested energy, which is a random variable (RV), is insufficient, a random number of REs may beamform in the reflection mode rather than a fixed number, with the rest blindly reflecting the incident signals. Thus, to fully understand the potential benefits and challenges of this integration, it is important to {efficiently optimize the REs allocation and develop a comprehensive analytical framework} to enhance and assess both the EH and communication performance. The paper's key contributions are as follows.
\begin{itemize}
    \item We consider a passive wireless system in which noise modulation is employed to generate information bits through unbiased resistors and convey them to the Rx via both direct and {zero-energy} RIS links in the presence of interference signals. The RIS REs are divided into two groups: one for {harvesting energy} from interference signals to enable the RIS beamforming functionality, and the other for reflecting the desired signals towards the Rx.     
     \item By considering a random number of REs capable of beamforming, depending on the harvested energy, and employing the time diversity technique of repetition coding, we derive analytical expressions for EH and communication performance metrics, such as energy constraint success rate (ECSR), BER, mutual information (MI), {EE}, and optimal threshold selection { under linear and nonlinear EH (LEH and NLEH) models.}
     \item {To jointly optimize the EH and communication performance,} we present a closed-form solution and efficient search algorithm for the RIS REs allocation to EH and signal reflection.
     \item {Since we consider the interference signals as potentially useful energy sources for enabling the RIS beamforming functionality, it is imperative to assess whether these signals are always beneficial for both the EH and communication performance. Thus,} we study the asymptotic behavior of the system in the high-interference regime. 
     \end{itemize}\vspace{-5pt}
     Extensive numerical and simulation results are provided to validate {the developed analytical results} and the proposed algorithm, as well as to demonstrate the superiority of the considered integration compared to conventional systems. The main findings of this paper are as follows.
     \vspace{-5pt}\begin{itemize}
     \item {{In both LEH and NLEH models,} by maximizing the ECSR through the proposed REs allocation algorithm, the communication performance is also optimized, as all REs in the reflection mode can effectively beamform the desired signals. However, if the number of REs allocated for EH exceeds the optimal level, the communication performance is degraded, as fewer REs remain for beamforming, making the extra harvested energy ineffective.}
     \item {It is observed that in low-interference regimes, the RIS beamforming capability is limited due to insufficient harvested energy, leading to insignificant performance gain. As interference increases, more energy can be harvested with fewer REs for EH, allowing more REs for beamforming, improving the communication performance. However, in high-interference regimes, while sufficient energy can be harvested with fewer REs for EH, the communication performance becomes severely restricted due to direct interference links. This highlights the need for interference cancellation in such environments, a promising area for future research.}
     \item {The proposed RIS RE allocation algorithm for EH and communication tasks is also shown to be significantly more efficient in terms of computational complexity compared to exhaustive search and random search methods, making it particularly well-suited for scenarios involving a large number of REs.}
     \item {It is demonstrated that the threshold selection and the resulting BER improvement depend on the numbers of REs and samples of the received signal. This is because increasing the REs used for beamforming raises the received signal strength, and a greater number of samples provides a more reliable estimate of the received signal's energy, enhancing the detection performance, which is further improved by employing the repetition coding technique. }
     \item {The results further highlight the crucial role of RISs in extending the inherently limited Tx–Rx communication range of conventional noise modulation-based systems. In addition, the proposed integration under both LEH and NLEH models demonstrates significantly superior EE performance compared to conventional RIS-assisted systems.}   
\end{itemize}
\vspace{-10pt}\begin{figure}[t]
\centering\includegraphics[width=0.55\linewidth]{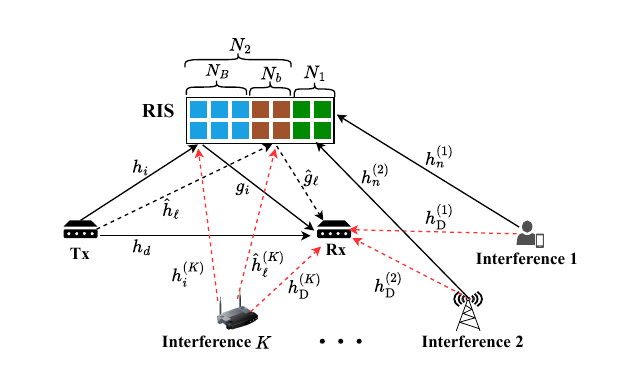}\vspace{-20pt}\caption{\ {{Zero-energy} RIS-assisted communications in the presence of interference. } }\label{sys}\vspace{-25pt}
\end{figure}
\subsection{Paper Organization and Mathematical Notation}\vspace{-5pt}
The system model, and the analytical expressions along with the optimization algorithm are presented in Sections \ref{SECII} and \ref{SECIII}, respectively. The results and conclusion of the paper are presented in Sections \ref{SECREsu} and \ref{SECCon}, respectively. Moreover, the mathematical notations are presented as follows. 

$X\sim\mathcal{CN}\left(\mu_{X},\sigma_{X}^{2}\right)$ denotes a circularly symmetric complex Gaussian RV with a mean and variance of $\mu_{X}$ and $\sigma_{X}^{2}$, respectively, and $X\sim U(a,b)$ denotes a uniformly distributed RV in the interval of $[a,b]$. Moreover, $\mu_{X}=\mathbb{E}[X]$ , $\sigma_{X}^{2}=\mathbb{V}[X]$, $|X|$, $\mathrm{P_{r}}(X)$, $\mathrm{P_{r}}(X|Y)$, and $\mathcal{M}_{X}(\mathcal{S})$ denote the expectation of $X$, variance of $X$, absolute value of $X$, probability of $X$, probability of $X$ conditioned on $Y$, and the moment generating function (MGF) of $X$, respectively. Additionally, $\Gamma(.)$ is the gamma function \cite[eq.~(8.310)]{bk}, $\Gamma(.,.)$ is the upper incomplete gamma function \cite[eq.~(8.350.2)]{bk}, $J_{n}(X)$ is the Bessel function of the first kind with an order of $n$ \cite[eq.~(8.402)]{bk}, $\lfloor X \rfloor$ is the floor function, $\lceil X \rceil$ is the ceiling function, $I\left(X;Y\right)$ is the MI between $X$ and $Y$, $H\left(Y\right)$ the marginal entropy, and $H\left(Y|X\right)$ is the conditional entropy. 
\vspace{-15pt}\section{System Model}\label{SECII}
 \vspace{-10pt}
As shown in Fig. \ref{sys}, we consider a wireless communication system, where a zero-energy Tx communicates with an Rx via both the direct link and indirect links through a zero-energy RIS. To {support a zero-energy} RIS, the REs are divided into two groups: (i) a group of $N_{1}$ REs operating in {the} energy absorption mode for harvesting energy from available RF signals of $K$ interferers, and (ii) a group of $N_{2}$ REs operating in {the} reflection mode for reflecting the incident signals towards the Rx, such that the total number of REs is $N=N_{1}+N_{2}$. Let $s_{k}$ represent the complex Gaussian signal of the $k$-th interferer, and thus {using the LEH model \cite[Eq.~(7)]{7999248},} the energy harvested by $N_{1}$ REs\footnote{Since the noise modulation-based signals transmitted by the Tx are inherently weak\cite{kapetanovic2022communication,9980386}, the harvested energy from them is neglected for simplicity. However, the ECSR expressions given in (\ref{pofeh}) and (\ref{pofehNL}) remain valid even if this energy is considered.} in a time slot with a duration of $t$ seconds is expressed as 
\begin{align}
    {\mathcal{E}_{H}^{\text{L}}= t\eta P_{H}}, \ { \text{with} \ \ P_{H} \triangleq \textstyle\sum_{k=1}^{K} P_{k}(d_{1}^{(k)})^{-a}\left|\textstyle\sum_{n=1}^{N_{1}}\tilde{h}_{n}^{(k)}e^{j\tilde{\theta}_{n}}\right|^{2}}\label{har},
\end{align}
where $\mathbb{E}[|s_{{k}}|^{2}]=P_{k}$ is the interference power, $d_{1}^{(k)}$ is the RIS$\to k$-th interferer distance, $a$ is the path-loss exponent,  $\tilde{h}_{n}^{(k)}$ is the fading channel described in TABLE \ref{TAB2}, and $\eta$ is the energy conversion efficiency. Moreover, as the CSI of the interference channels is unknown, the phase shifts of the $N_{1}$ REs are assumed to follow {the} uniform distribution as $\tilde{\theta}_{n}\sim U(0,2\pi)$ \cite{10356096,9440664}. 
\begin{table}[t]
  \caption{\ {Notations for Fading Channels and RIS Phase Shifts}}\vspace{-20pt}\centering
 \begin{center}\begin{tabular}{|p{0.21\linewidth}|p{0.724\linewidth}|}\hline
\cellcolor[gray]{.75}{\textbf{Symbol}} & \cellcolor[gray]{.75}{\textbf{Description}} \\ \hline
{$\tilde{h}_{n}^{(k)}\sim\mathcal{CN}\left(0,1\right)$} & {Fading channel coefficient between the $k$-th interferer and the $n$-th RE, $n\in\{1,\dots,N_{1}\}$} \\ \hline
{$\tilde{\theta}_{n}\in[0,2\pi)$} & {Phase shift applied by the $n$-th RE} \\ \hline
{$h_{\text{d}}\sim\mathcal{CN}\left(\mu_{\text{d}},1\right)$} & {Fading channel coefficient between the Tx and Rx} \\ \hline
{$h_{i}\sim\mathcal{CN}\left(0,1\right)$} & {Fading channel coefficient between the Tx and the $i$-th RE, $i\in\{1,\dots,N_{B}\}$} \\ \hline
{$g_{i}\sim\mathcal{CN}\left(0,1\right)$} & {Fading channel coefficient between the $i$-th RE and Rx } \\ \hline
{$\theta_{i}\in[0,2\pi)$} & {Phase shift applied by the $i$-th RE}  \\ \hline
{$\phi_{i}\in[-\pi/Q,\pi/Q]$} & {Phase shift quantization error of the $i$-th RE} \\ \hline
{$\hat{h}_{\ell}\sim\mathcal{CN}\left(0,1\right) $} & {Fading channel coefficient between the Tx and the $\ell$-th RE, $\ell\in\{1,\dots,N_{b}\}$} \\ \hline
{$\hat{g}_{\ell}\sim\mathcal{CN}\left(0,1\right)$ } & {Fading channel coefficient between the $\ell$-th RE and Rx } \\ \hline
{$\hat{\theta}_{\ell}=0$} & {Phase shift of the $\ell$-th RE} \\ \hline
{$h_{\text{D}}^{(k)}\sim\mathcal{CN}\left(0,1\right)$ } & {Fading channel coefficient between the $k$-th interferer and Rx} \\ \hline
{$h_{i}^{(k)}\sim\mathcal{CN}\left(0,1\right)$ } & {Fading channel coefficient between the $k$-th interferer and the $i$-th RE} \\ \hline
{$\hat{h}_{\ell}^{(k)}\sim\mathcal{CN}\left(0,1\right)$ } & {Fading channel coefficient between $k$-th interferer and the $\ell$-th RE} \\ \hline
 \end{tabular}
 \end{center}\vspace{-40pt}
\label{TAB2}
\end{table}

\vspace{-20pt}{It is noteworthy that, since the energy conversion efficiency in the LEH model is constant and independent of the input power, it does not account for the non-linear distortions or saturation that arise in practical EH systems \cite{7999248,10778216}. For practicality and comparison purposes, we adopt the NLEH model proposed and validated through experimental data in \cite{7999248}. Thus, the energy harvested by $N_{1}$ REs via the NLEH model can be expressed as \cite[Eq.~(9)]{7999248}}
\begin{align}
    {\mathcal{E}_{H}^{\text{NL}}=t\left(\frac{\mathcal{A}_{1}P_{H}+\mathcal{A}_{2}}{P_{H}+\mathcal{A}_{3}}-\frac{\mathcal{A}_{2}}{\mathcal{A}_{3}}\right) \ ,}\label{NLEH1}
\end{align}
{where $P_{H}$ is given in (\ref{har}), and $\{\mathcal{A}_{1}, \mathcal{A}_{2}, \mathcal{A}_{3}\}=\{2.463, 1.635, 0.826\}$ are constants specified by the standard curve-fitting \cite{7999248,10778216}.}  

To enable the RIS beamforming, it is essential to quantify the energy consumption of the $N_{2}$ REs and the RIS controller along with the EH circuitry, which can be expressed as \cite{10348506,9866567} 
\begin{align}
    \mathcal{E}_{C}=t\left(N_{2}\mathcal{E}+\mathcal{E}_{con}\right),\label{con}
\end{align}
where $\mathcal{E}$ is the power consumption of each RE, and $\mathcal{E}_{con}$ is the power consumption of the RIS controller along with the EH circuitry. 

 To ensure that all the $N_{2}$ REs can adopt appropriate phase shifts for beamforming, the energy constraint, {$\mathcal{E}_{H}^{\zeta}\geq \mathcal{E}_{C}$ where $\zeta\in\{\text{L},\text{NL}\}$}, must be satisfied. Since {$\mathcal{E}_{H}^{\zeta}$} is a RV, the number of REs that are capable of beamforming the desired signals also becomes a RV. When ${\mathcal{E}_{H}^{\zeta}}<\mathcal{E}_{C}$, a subset of the $N_{2}$ REs, denoted by $N_{B}$, is capable of beamforming and applying phase shifts, i.e., $\theta_{i}\in[0,2\pi)$ with $i\in\{1,\dots, N_{B}\}$, while the remaining REs, denoted by $N_{b}$, blindly reflect the signals without applying phase shifts, i.e., $\hat{\theta}_{\ell}=0$ with $\ell\in\{1,\dots, N_{b}\}$\cite{9034159,10557632,8801961}, such that $N_{2}=N_{B}+N_{b}$. Therefore, $N_{B}$ and $N_{b}$ are binomial RVs with $\mathbb{E}[N_{B}]=N_{2}{\mathcal{P}_{s}^{\zeta}}$, $\mathbb{E}[N_{b}]=N_{2}(1-{\mathcal{P}_{s}^{\zeta}})$, and $\mathbb{V}[N_{B}]=\mathbb{V}[N_{b}]=N_{2}{\mathcal{P}_{s}^{\zeta}}(1-{\mathcal{P}_{s}^{\zeta}})$\cite{ROSS}, where ${\mathcal{P}_{s}^{\zeta}}=\mathrm{P_{r}}\left({\mathcal{E}_{H}^{\zeta}}\geq \mathcal{E}_{C}\right)$ is the {ECSR - the} probability of satisfying the energy constraint. 
\begin{figure}[t!]
\centering\includegraphics[width=\linewidth]{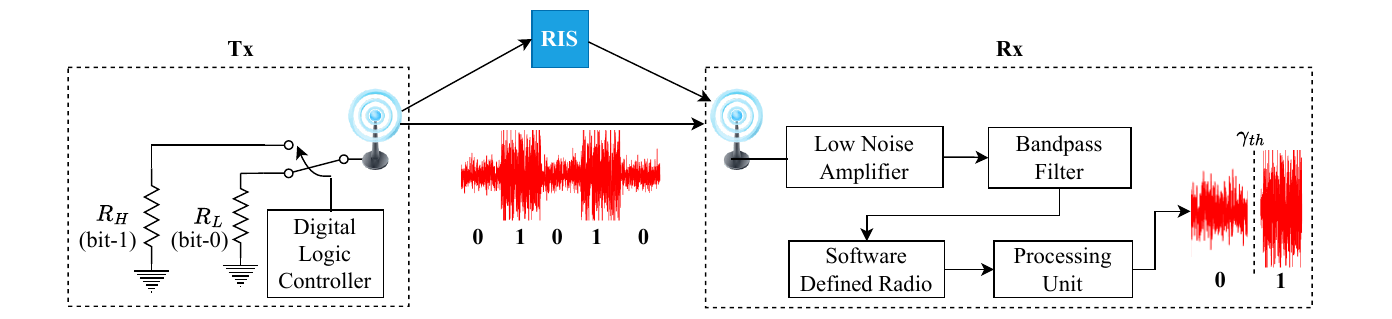}\vspace{-10pt}\caption{\ The generic transceiver architecture of RIS-assisted thermal noise modulation-based communications. }\label{archt}\vspace{-25pt}
\end{figure}

{In order to achieve zero-energy transmission,} the thermal noise modulation scheme is employed, where its transceiver architecture is shown in Fig. \ref{archt}. In this modulation scheme, bit-0 and bit-1 waveforms are generated by adjusting the noise power sources, such as a low-value resistor, $R_{L}$, and a high-value resistor, $R_{H}$, without an external power source, where a digital logic controller is used to control switching between the two resistors\cite{kapetanovic2022communication,9980386,kapetanovic2023cosmic,10373568}. The waveforms are then transmitted to the Rx via both the direct and RIS links. At the Rx end, the signal is boosted by a low noise amplifier, and a bandpass filter is used to ensure that the Rx only processes the part of the noise spectrum that carries the information\cite{kapetanovic2022communication}. {Afterwards}, the software-defined radio (SDR) and processing unit are used to capture and downconvert the signal to baseband samples, {and to perform bit detection by comparing the energy of the received signal to a decision threshold \cite{kapetanovic2022communication,10373568,9980386}.} 

To further enhance the performance, the time diversity technique {via} repetition coding with $\mathcal{R}$ repetitions is used, which is suitable for low-power applications\cite{10373568,31130,4539783}. In this approach, $M$ independent and identically distributed (IID) samples are received in each repetition\cite{31130,4539783}. In the $r$-th repetition, the $m$-th sample of the zero-mean complex Gaussian transmit signal is denoted as $s_{m,r}$ with variances of $\sigma_{0}^{2}=4\kappa T R_{L}B$ {if} $s_{m,r}$ is bit-0 and $\sigma_{1}^{2}=4\kappa T R_{H}B\triangleq\mathcal{C}\sigma_{0}^{2}$ {if} $s_{m,r}$ is bit-1, where $\kappa$ is the Boltzmann’s constant, $T$ is the temperature in Kelvin, $B$ is the bandwidth, and $\mathcal{C}$ is a coefficient specifying the relationship between the two variances and also the resistors as $R_{H}=\mathcal{C} R_{L}$, respectively \cite{kapetanovic2022communication,10373568,10706852}. Moreover, $K$ interference signals also arrive at the Rx through both direct links and indirect links provided by the $N_{2}$ REs. Thus, in the $r$-th repetition, the $m$-th sample of the received signal at the Rx is expressed as
\vspace{-5pt}\begin{align}
x_{m,r}=&\underbrace{\Big(\mathcal{L}_{\text{d}}h_{\text{d}}+\mathcal{L}_{B}\sum_{i=1}^{N_{B}}h_{i}g_{i}e^{j\theta_{i}}+\mathcal{L}_{B}\sum_{\ell=1}^{N_{b}}\hat{h}_{\ell}\hat{g}_{\ell}\Big)s_{m,r}}_{\text{Desired signal}}\nonumber\\
&+\underbrace{\sum_{k=1}^{K}\Big(\mathcal{L}_{\text{D}}^{(k)}h_{\text{D}}^{(k)}+\mathcal{L}_{B}^{(k)}\sum_{i=1}^{N_{B}}h_{i}^{(k)}g_{i}e^{j\theta_{i}}+\mathcal{L}_{B}^{(k)}\sum_{\ell=1}^{N_{b}}\hat{h}_{\ell}^{(k)}\hat{g}_{\ell}\Big)s_{k,m,r}}_{\text{Interference signals}}+w_{m,r},\label{rec1}
\end{align}
where $w_{m,r}\sim\mathcal{CN}(0,N_{0})$ is the AWGN, and $s_{k,m,r}$ with a variance of $P_{k}$ is the $k$-th interferer signal. Moreover, $\mathcal{L}_{\text{d}}=d^{-a/2}$, $\mathcal{L}_{\text{D}}^{(k)}=(d_{\text{D}}^{(k)})^{-a/2}$, $\mathcal{L}_{B}=(d_{1}d_{2})^{-a/2}$, and $\mathcal{L}_{B}^{(k)}=({d}_{1}^{(k)}d_{2})^{-a/2}$ are the path-losses, where $d$, $d_{\text{D}}^{(k)}$, $d_{1}$, and $d_{2}$ denote the Tx$\to$Rx distance, the Rx$\to k$-th interferer distance, the Tx$\to$RIS distance, and the RIS$\to$Rx distance, respectively. The remaining notations are given in TABLE \ref{TAB2}. 

\vspace{-20pt}\section{Performance Analysis}\label{SECIII}
In this section, we derive analytical expressions for the ECSR, BER, optimal threshold, MI, and EE. Moreover, we present an {approximated} closed-form solution and an optimization algorithm for the optimal RIS resource allocation. Additionally, we provide a further simplified lower-bound on the BER and analyze the asymptotic behavior of the system performance.
\vspace{-10pt}\subsection{Energy Constraint Success Rate (ECSR)}\label{sec3a}
{We first consider the LEH model.} By exploiting (\ref{har}) and (\ref{con}), the ECSR, ${\mathcal{P}_{s}^{\text{L}}=\mathrm{P_{r}}\left(\mathcal{E}_{H}^{\text{L}}\geq \mathcal{E}_{C}\right)}$, is expressed as
\vspace{-10pt}\begin{align}
    {\mathcal{P}_{s}^{\text{L}}=\mathrm{P_{r}}\left(t\eta P_{H}\geq t\left(N_{2}\mathcal{E}+\mathcal{E}_{con}\right)\right)=\mathrm{P_{r}}\left( P_{H}\geq \frac{N_{2}\mathcal{E}+\mathcal{E}_{con}}{\eta}\right)}.\label{ECSRmain}
\end{align}

Since $\tilde{h}_{n}^{(k)}$ is a complex Gaussian RV and $\tilde{\theta}_{n}\sim U(0,2\pi)$ is a uniformly distributed RV, by virtue of \cite[Proposition 2]{9440664} and \cite[Sec. III-B]{10047976}, it can be {established} that $x=\sum_{n=1}^{N_{1}}\tilde{h}_{n}^{(k)}e^{j\tilde{\theta}_{n}}\sim\mathcal{CN}\left(0, N_{1} \right)$. {Therefore}, $|x|^{2}$ follows the exponential distribution, {and $ P_{H}$ given in (\ref{har}) is} the scaled sum of exponential RVs, which can be modelled by the gamma distribution. Thus, by exploiting\cite[TABLE~ 2.2]{ROSS}, ${\mathcal{P}_{s}^{\text{L}}}$ is obtained as 
\vspace{-10pt}\begin{align}
    {\mathcal{P}_{s}^{\text{L}}}=\frac{1}{\Gamma(\alpha)}\Gamma\left(\alpha,\frac{\beta \left(N_{2}\mathcal{E}+\mathcal{E}_{con}\right)}{\eta}\right),\label{pofeh}
\end{align}
where $\alpha=\frac{\mu_{P_{H}}^{2}}{\sigma_{P_{H}}^{2}}$, $\beta=\frac{\mu_{P_{H}}}{\sigma_{P_{H}}^{2}}$, $\mu_{P_{H}}=\sum_{k=1}^{K} P_{k}(d_{1}^{(k)})^{-a}N_{1}$ and $\sigma_{P_{H}}^{2}=\sum_{k=1}^{K}\left( P_{k}(d_{1}^{(k)})^{-a}N_{1}\right)^{2}$.

{Similarly, for the NLEH model, by exploiting (\ref{NLEH1}) and (\ref{con}), the ECSR, $\mathcal{P}_{s}^{\text{NL}}=\mathrm{P_{r}}\left(\mathcal{E}_{H}^{\text{NL}}\geq \mathcal{E}_{C}\right)$, is expressed as}
\begin{align}
    {\mathcal{P}_{s}^{\text{NL}}}&{=\mathrm{P_{r}}\left(\frac{\mathcal{A}_{1}P_{H}+\mathcal{A}_{2}}{P_{H}+\mathcal{A}_{3}}-\frac{\mathcal{A}_{2}}{\mathcal{A}_{3}}\geq \left(N_{2}\mathcal{E}+\mathcal{E}_{con}\right)\right)}\nonumber\\
    &{=\mathrm{P_{r}}\left(P_{H}\left(\mathcal{A}_{1}-N_{2}\mathcal{E}-\mathcal{E}_{con}-\frac{\mathcal{A}_{2}}{\mathcal{A}_{3}}\right)\geq \mathcal{A}_{3}\left(N_{2}\mathcal{E}+\mathcal{E}_{con}\right)\right)}\nonumber\\
    &{=\mathrm{P_{r}}\left(P_{H}\geq \frac{\mathcal{A}_{3}^{2}\left(N_{2}\mathcal{E}+\mathcal{E}_{con}\right)}{\mathcal{A}_{1}\mathcal{A}_{3}-\mathcal{A}_{3}\left(N_{2}\mathcal{E}+\mathcal{E}_{con}\right)-\mathcal{A}_{2}}\right).}\label{ECNL1}
\end{align}

{As discussed above, since $P_{H}$ can be modelled by the gamma distribution, (\ref{ECNL1}) is written as}
\begin{align}
    {\mathcal{P}_{s}^{\text{NL}}=\frac{1}{\Gamma(\alpha)}\Gamma\left(\alpha,\frac{\beta\mathcal{A}_{3}^{2}\left(N_{2}\mathcal{E}+\mathcal{E}_{con}\right)}{\mathcal{A}_{1}\mathcal{A}_{3}-\mathcal{A}_{3}\left(N_{2}\mathcal{E}+\mathcal{E}_{con}\right)-\mathcal{A}_{2}}\right),}\label{pofehNL}
\end{align}
{where $\alpha$ and $\beta$ are the same as those defined in (\ref{pofeh}).} 
\vspace{-15pt}\subsection{RIS Resource Allocation}\vspace{-5pt}
As discussed earlier, to ensure that all the $N_{2}$ REs can beamform the signals, ${\mathcal{E}_{H}^{\zeta}}\geq \mathcal{E}_{C}$ is required. Thus, for an efficient RIS resource allocation, we aim to minimize the number of REs operating in {the} absorption mode, $N_{1}$, which leads to the maximization of $N_{2}$ and ${\mathcal{P}_{s}^{\zeta}}=\mathrm{P_{r}}\left({\mathcal{E}_{H}^{\zeta}}\geq \mathcal{E}_{C}\right)$. The corresponding optimization problem is formulated as
\vspace{-10pt}\begin{align}
    &\text{min}\ \  N_{1}=N-N_{2}\nonumber\\
     &\text{s.t.}\ \ {\mathcal{E}_{H}^{\zeta}}\geq \mathcal{E}_{C}, \ N_{1}\geq 1, \ N_{2}\geq 1
\end{align}

{Since the harvested energy increases as $N_1$ increases, it can be stated that ${\mathcal{P}_{s}^{\zeta}}$ increases monotonically with $N_1$. Consequently, if ${\mathcal{P}_{s}^{\zeta}} = 1$ holds for a particular $N_1$, it will also hold for any larger value of $N_1$. Conversely, if ${\mathcal{P}_{s}^{\zeta}} < 1$ for a given $N_1$, the condition will remain ${\mathcal{P}_{s}^{\zeta}} < 1$ for any smaller value of $N_1$. This monotonicity implies that it is a quasi-convex problem, as there exists a unique smallest $N_1$ that satisfies the condition \cite{convex}. Hence, the binary search is an appropriate method for solving this problem, as it reduces the search range by halving the interval in each iteration \cite{binsearch}. Additionally, this method significantly reduces the computational complexity compared to the exhaustive search method, which requires evaluating ${\mathcal{P}_{s}^{\zeta}}$ for all possible values of $N_1$ \cite{binsearch}. {Therefore, the binary search algorithm presented in Algorithm \ref{ALG1}, whose flow‐process diagram is illustrated in Fig. \ref{flowdiag}, provides an efficient and optimal solution to the aforementioned minimization problem.}
\vspace{-10pt}\begin{algorithm}
 \caption{Binary Search Algorithm for RIS RE Allocation}\label{ALG1}
 \begin{algorithmic}[1]
 \renewcommand{\algorithmicrequire}{\textbf{Input:}}
 \renewcommand{\algorithmicensure}{\textbf{Output:}}
 \REQUIRE $N$, $K$, $\eta$, ${\mathcal{A}_{1}}$, ${\mathcal{A}_{2}}$, ${\mathcal{A}_{3}}$, $P_{k}$, $d_{1}^{(k)}$, $a$, $\mathcal{E}$, $\mathcal{E}_{con}$
 \ENSURE  $ N_{1}^{*}$, $ N_{2}^{*}=N-N_{1}^{*}$
 \\ \textit{Initialization} :
  \STATE Set the lower-bound and upper-bound on the number of REs for EH as $N_{1,\text{L}}\leftarrow 1$ and $N_{1,\text{U}}\leftarrow (N-1)$
 \\ \textit{Search Process:}
  \WHILE {$N_{1,\text{L}}\leq N_{1,\text{U}}$}
  \STATE Calculate the midpoint, i.e., $N_{1}=\left\lfloor\frac{N_{1,\text{L}}+ N_{1,\text{U}}}{2}\right\rfloor$
  \STATE $N_{2}\leftarrow (N-N_{1})$
  \STATE Calculate ${\mathcal{P}_{s}^{\zeta}}$ using (\ref{pofeh}) {for LEH model and (\ref{pofehNL}) for NLEH model}
  \IF{${\mathcal{P}_{s}^{\zeta}}=\mathrm{P_{r}}\left({\mathcal{E}_{H}^{\zeta}}\geq \mathcal{E}_{C}\right)=1$}
  \STATE Update $N_{1,\text{U}}\leftarrow (N_1-\mathbf{1}(N_1=N_{1,\text{U}}))$, where $\mathbf{1}$ is the indicator function
  \ELSIF{${\mathcal{P}_{s}^{\zeta}}< 1$ \AND $N_{1} = N-1$}
  \STATE Break and return $N_{1}^{*} = N-1$ and $N_{2}^{*} = 1$
  \ELSE 
  \STATE Update $N_{1,\text{L}}\leftarrow (N_1+\mathbf{1}(N_1=N_{1,\text{L}}))$
  \ENDIF
  \ENDWHILE
 \RETURN $N_{1}^{*}$, which is the smallest value of $N_{1}$ that satisfies ${\mathcal{P}_{s}^{\zeta}}=\mathrm{P_{r}}\left({\mathcal{E}_{H}^{\zeta}}\geq \mathcal{E}_{C}\right)=1$ and leads to the maximum number of REs capable of beamforming, or $N_{1}^{*}=N-1$ and $N_{2}^{*}=1$ if no such $N_{1}$ exists.
 \end{algorithmic} 
 \end{algorithm}
\vspace{-30pt}\begin{figure}[hb!]
\centering\includegraphics[width=0.85\linewidth]{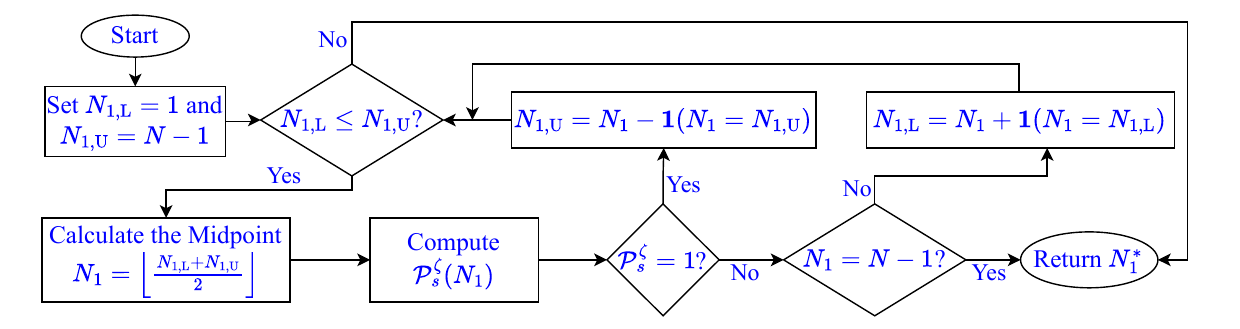}\vspace{-10pt}\caption{\ {Flow-process diagram of the RIS RE allocation.}}\label{flowdiag}
\end{figure}

To gain insights, considering the average harvested energy and using (\ref{har}), an approximated closed-form solution for the RIS REs allocation, $N_{1}^{*}$, can be derived as
\vspace{-10pt}\begin{align}
{\eta\mathbb{E}[P_{H}]\geq \left(N_{2}\mathcal{E}+\mathcal{E}_{con}\right)}\nonumber\\
    \eta\textstyle\sum_{k=1}^{K} P_{k}(d_{1}^{(k)})^{-a}\mathbb{E}\Big[\Big|\textstyle\sum_{n=1}^{N_{1}}\tilde{h}_{n}^{(k)}e^{j\tilde{\theta}_{n}}\Big|^{2}\Big]\geq \left(N_{2}\mathcal{E}+\mathcal{E}_{con}\right).\label{close1}
\end{align}

{As explained in Sec. \ref{sec3a}, based on \cite[Proposition 2]{9440664} and \cite[Sec. III-B]{10047976}, it can be concluded that $x\triangleq\sum_{n=1}^{N_{1}}\tilde{h}_{n}^{(k)}e^{j\tilde{\theta}_{n}}$ converges to a complex Gaussian RV as $x\sim\mathcal{CN}\left(0, N_{1} \right)$, which results in $\mathbb{E}\left[|x|^{2}\right]=\mathbb{E}\Big[\left|\sum_{n=1}^{N_{1}}\tilde{h}_{n}^{(k)}e^{j\tilde{\theta}_{n}}\right|^{2}\Big]=N_{1}$. Therefore, (\ref{close1}) can be rewritten as}
\begin{align}
    \textstyle\sum_{k=1}^{K}\eta P_{k}(d_{1}^{(k)})^{-a}N_{1}\geq \left(N\mathcal{E}-N_{1}\mathcal{E}+\mathcal{E}_{con}\right)\nonumber\\
    N_{1}^{*}\geq\Big\lceil\frac{N\mathcal{E}+\mathcal{E}_{con}}{\eta\sum_{k=1}^{K} P_{k}(d_{1}^{(k)})^{-a}+\mathcal{E}}\Big\rceil.\label{close}
\end{align}
\newtheorem{Remark}{Remark}\begin{Remark}\label{rem1}From (\ref{close}), it is evident that {the number of REs required for EH is a monotonically decreasing function of the number of interferers.}
\end{Remark} 
\vspace{-27pt}\subsection{Accurate Bit Error Rate (BER) Analysis}\label{SECII-C}
\vspace{-8pt}Taking into account the repetition coding technique, the BER of the system with $\mathcal{R}$ repetitions can be obtained by applying a simple majority vote decoding method {\cite{31130,4539783}}. In this method, the Rx decodes the bit as 1 if more 1s than 0s are received, {otherwise a bit of 0 is declared} \cite{31130,4539783}. Therefore, the BER of the system with $\mathcal{R}$ repetitions can be expressed as\cite[Eq.~(2)]{31130}
\vspace{-5pt}\begin{align}
    \widehat{\mathcal{P}}_{b}=\sum_{r=(\mathcal{R}+1)/2}^{\mathcal{R}}\frac{\mathcal{R}!}{r!(\mathcal{R}-r)!}\mathcal{P}_{b}^{r}\left(1-\mathcal{P}_{b}\right)^{\mathcal{R}-r},\label{ABERsys}
\end{align}
where $\mathcal{P}_{b}$ is the BER in the $r$-th repetition, and $\mathcal{R}$ is odd. 

To derive $\mathcal{P}_{b}$, the energies of the received signal over $M$ samples for both bit-0 and bit-1 in the $r$-th repetition are compared with a threshold, $\gamma_{th}$, which is expressed as\cite{10269022}
\vspace{-5pt}\begin{align}
    \mathcal{P}_{b}=\frac{1}{2}\mathrm{P_{r}}\left(\textstyle\sum_{m=1}^{M}\left|x_{m,r}\right|^{2}>\gamma_{th}|\text{bit-0}\right)+\frac{1}{2}\mathrm{P_{r}}\left(\textstyle\sum_{m=1}^{M}\left|x_{m,r}\right|^{2}\leq \gamma_{th}|\text{bit-1}\right).\label{Pbr}
\end{align}

Since $s_{m,r}$, $s_{k}$, and $w_{m,r}$ are zero-mean complex Gaussian RVs, the received signal given in (\ref{rec1}) is also a complex Gaussian RV but conditioned on the random number of REs, fading channels and RIS phase shifts, such that $x_{m,r}\sim\mathcal{CN}(0,\sigma_{x,0}^{2})$ for bit-0 and $x_{m,r}\sim\mathcal{CN}(0,\sigma_{x,1}^{2})$ for bit-1, where the conditional variances, $\sigma_{x,\chi}^{2} \ \chi\in\{0,1\}$, can be written as   
\vspace{-5pt}\begin{align}
   \sigma_{x,\chi}^{2}=
    \begin{cases}
     \sigma_{x,0}^{2}=\mathcal{D}\sigma_{0}^{2}+\sum_{k=1}^{K}P_{k}\mathcal{I}_{k}+N_{0},&  \text{for bit-0}\\ 
    \sigma_{x,1}^{2}=\mathcal{D}\sigma_{1}^{2}+\sum_{k=1}^{K}P_{k}\mathcal{I}_{k}+N_{0},&  \text{for bit-1},\label{convar}
\end{cases}
\end{align}\vspace{-15pt}
where 
\vspace{-10pt}\begin{align}
\mathcal{D}=\big|\mathcal{L}_{\text{d}}h_{\text{d}}+\mathcal{L}_{B}\textstyle\sum_{i=1}^{N_{B}}h_{i}g_{i}e^{j\theta_{i}}+\mathcal{L}_{B}\textstyle\sum_{\ell=1}^{N_{b}}\hat{h}_{\ell}\hat{g}_{\ell}\big|^{2}.\label{D1}
\end{align}
\vspace{-15pt}\begin{align}
\mathcal{I}_{k}=\big|\mathcal{L}_{\text{D}}^{(k)}h_{\text{D}}^{(k)}+\mathcal{L}_{B}^{(k)}\textstyle\sum_{i=1}^{N_{B}}h_{i}^{(k)}g_{i}e^{j\theta_{i}}+\mathcal{L}_{B}^{(k)}\textstyle\sum_{\ell=1}^{N_{b}}\hat{h}_{\ell}^{(k)}\hat{g}_{\ell}\big|^{2}.
\end{align}

\vspace{-5pt}Considering the conditionally complex Gaussian RV, $x_{m,r}$, it can be stated that $|x_{m,r}|^{2}$ follows a conditional exponential distribution, and thus $z=\sum_{m=1}^{M}\left|x_{m,r}\right|^{2}$ follows a conditional central Chi-squared distribution with $2M$ degrees of freedom\cite{ROSS}. Therefore, using the cumulative distribution function of the Chi-squared distribution given in \cite[Eq.~(2.3-24)]{THz42}, the conditional probabilities, $\mathcal{P}_{c|0}\triangleq \mathrm{P_{r}}\left(\sum_{m=1}^{M}\left|x_{m,r}\right|^{2}>\gamma_{th}|\text{bit-0}\right)$ and $\mathcal{P}_{c|1}\triangleq \mathrm{P_{r}}\left(\sum_{m=1}^{M}\left|x_{m,r}\right|^{2}\leq \gamma_{th}|\text{bit-1}\right)$ are respectively written as 
\vspace{-10pt}\begin{align}
    \mathcal{P}_{c|0}=\frac{1}{\Gamma(M)}\Gamma\left(M,\frac{\gamma_{th}}{\sigma_{x,0}^{2}}\right)=\frac{1}{\Gamma(M)}\Gamma\left(M,\frac{\gamma_{th}}{\mathcal{D}\sigma_{0}^{2}+\sum_{k=1}^{K}P_{k}\mathcal{I}_{k}+N_{0}}\right).\label{cb0}
\end{align}
\vspace{-10pt}\begin{align}
    \mathcal{P}_{c|1}=1-\frac{1}{\Gamma(M)}\Gamma\left(M,\frac{\gamma_{th}}{\sigma_{x,1}^{2}}\right)=1-\frac{1}{\Gamma(M)}\Gamma\left(M,\frac{\gamma_{th}}{\mathcal{D}\sigma_{1}^{2}+\sum_{k=1}^{K}P_{k}\mathcal{I}_{k}+N_{0}}\right).\label{cb1}
\end{align}

The unconditional probabilities, $\mathcal{P}_{0}=\mathbb{E}[\mathcal{P}_{c|0}]$ and $\mathcal{P}_{1}=\mathbb{E}[\mathcal{P}_{c|1}]$, are obtained by taking the expectations of (\ref{cb0}) and (\ref{cb1}) with respect to the conditional variances, $\sigma_{x,0}^{2}$ and $\sigma_{x,1}^{2}$. Since the RVs $\sigma_{x,0}^{2}$ and $\sigma_{x,1}^{2}$ depend on a large number of desired and interference fading channels, random phases, and the random number of REs, deriving accurate expressions for the unconditional probabilities via conventional methods requires at least $(K+1)$-fold numerical integration, which is mathematically {challenging} and burdensome\cite{10269022}. Alternatively, by leveraging an MGF-based method proposed in \cite[Lemma 1]{6425522}, the unconditional probabilities can be obtained as 
\begin{align}
    \mathcal{P}_{0}&=1-\frac{1}{\Gamma(M)}\int_{0}^{\infty}\mathcal{S}^{\frac{M}{2}-1}J_{M}\left(2\sqrt{\mathcal{S}}\right)\mathcal{M}_{\sigma_{x,0}^{2}}\left(\mathcal{S}\right)\text{d}\mathcal{S},\label{P0}
\end{align}
\begin{align}
    \mathcal{P}_{1}&=\frac{1}{\Gamma(M)}\int_{0}^{\infty}\mathcal{S}^{\frac{M}{2}-1}J_{M}\left(2\sqrt{\mathcal{S}}\right)\mathcal{M}_{\sigma_{x,1}^{2}}\left(\mathcal{S}\right)\text{d}\mathcal{S},\label{P1}
\end{align}
where $\mathcal{M}_{\sigma_{x,\chi}^{2}}\left(\mathcal{S}\right)=\mathbb{E}\left[e^{\frac{-\mathcal{S}}{\gamma_{th}}\sigma_{x,\chi}^{2}}\right]$ $\chi\in\{0,1\}$ represents the MGF, which can be expressed as
\begin{align}
    \mathcal{M}_{\sigma_{x,\chi}^{2}}\left(\mathcal{S}\right)&=\mathbb{E}\left[e^{-\frac{\mathcal{S}}{\gamma_{th}}\left(\mathcal{D}\sigma_{\chi}^{2}+\sum_{k=1}^{K}P_{k}\mathcal{I}_{k}+N_{0}\right)}\right]\nonumber\\
    &=e^{-\frac{\mathcal{S}N_{0}}{\gamma_{th}}}\mathbb{E}\left[e^{-\frac{\mathcal{S}}{\gamma_{th}}\mathcal{D}\sigma_{\chi}^{2}}\right]\mathbb{E}\left[e^{-\frac{\mathcal{S}}{\gamma_{th}}\sum_{k=1}^{K}P_{k}\mathcal{I}_{k}}\right]\nonumber\\
    &=e^{-\frac{\mathcal{S}N_{0}}{\gamma_{th}}}\mathcal{M}_{D,\chi}\left(\mathcal{S}\right)\textstyle\prod_{k=1}^{K}\mathcal{M}_{\mathcal{I}_{k}}\left(\mathcal{S}\right),\label{MGFmain}
\end{align}
where $\mathcal{M}_{D,\chi}\left(\mathcal{S}\right)=\mathbb{E}\left[e^{-\frac{\mathcal{S}}{\gamma_{th}}\mathcal{D}\sigma_{\chi}^{2}}\right]$ and $\mathcal{M}_{\mathcal{I}_{k}}\left(\mathcal{S}\right)=\mathbb{E}\left[e^{-\frac{\mathcal{S}}{\gamma_{th}}P_{k}\mathcal{I}_{k}}\right]$ are the MGFs of $\mathcal{D}$ and $\mathcal{I}_{k}$.  

To achieve ideal beamforming for the desired signals, based on the availability of CSI at the RIS, the phase shifts applied by $N_{B}$ REs, $\theta_{i}$, should be {adjusted} to align with the phases of the fading channels, such that $\theta_{i}=\angle{h_{\text{D}}}-(\angle h_{i}+ \angle g_{i})$\cite{9440664}. However, in practice the RIS REs are equipped with discrete phase shifters which are only able to apply quantized phase shifts from the set of uniformly quantized phases, $\{0, 2\pi/Q,\dots,2\pi(Q-1)/Q\}$, where $Q=2^b$ is the quantization level and $b$ is the number of quantization bits\cite{10047976,10328394}. Therefore, the desired signals are reflected towards the Rx with phase shift quantization errors denoted as $\phi_{i}$, which are uniformly distributed in the interval of $[-\pi/Q,\pi/Q]$\cite{10047976,10328394}. In light of the above, $\mathcal{D}$ given in (\ref{D1}) can be rewritten as
\begin{align}
\mathcal{D}=\left|\mathcal{L}_{\text{d}}|h_{\text{d}}|+\mathcal{L}_{B}\textstyle\sum_{i=1}^{N_{B}}|h_{i}||g_{i}|e^{j\phi_{i}}+\mathcal{L}_{B}\textstyle\sum_{\ell=1}^{N_{b}}\hat{h}_{\ell}\hat{g}_{\ell}\right|^{2}.\label{Dp}
\end{align}

Since $\mathcal{D}$ involves the sums of positive RVs, by virtue of \cite{10328394,10057425,10316584}, it can be accurately modelled by {the} gamma distribution, where its MGF is expressed as\cite[Eq.~(27)]{10328394}
\begin{align}
    \mathcal{M}_{D,\chi}\left(\mathcal{S}\right)=\left(1+\frac{\sigma_{\chi}^{2}\mathcal{S}\sigma_{\mathcal{D}}^{2}}{\gamma_{th}\mu_{\mathcal{D}} }\right)^{-\frac{\mu_{\mathcal{D}}^{2}}{\sigma_{\mathcal{D}}^{2}}},\label{MGFD}
\end{align}
where $\mu_{\mathcal{D}}$ and $\sigma_{\mathcal{D}}^{2}$ are the mean and variance of $\mathcal{D}$, {which can be respectively} expressed as
\begin{align}
\mu_{\mathcal{D}}=\mathbb{V}\left[\mathcal{D}_{\text{R}}\right]+\mathbb{V}\left[\mathcal{D}_{\text{I}}\right]+\mathbb{E}^{2}\left[\mathcal{D}_{\text{R}}\right],\label{MEAND}
\end{align}
\begin{align}
    \sigma_{\mathcal{D}}^{2}=2\mathbb{V}\left[\mathcal{D}_{\text{R}}\right](\mathbb{V}\left[\mathcal{D}_{\text{R}}\right]+2\mathbb{E}^{2}\left[\mathcal{D}_{\text{R}}\right])+2\mathbb{V}^{2}\left[\mathcal{D}_{\text{I}}\right],\label{VARD}
\end{align}
where
\vspace{-10pt}\begin{align}
\mathbb{E}\left[\mathcal{D}_{\text{R}}\right]=\frac{\mathcal{L}_{\text{d}}\sqrt{\pi}}{2}L_{\frac{1}{2}}\left(-\mu_{\text{d}}^{2}\right)+\frac{\mathcal{L}_{B}N_{2}{\mathcal{P}_{s}^{\zeta}}Q}{4}\sin\left(\frac{\pi}{Q}\right),
\end{align}
\begin{align}
\mathbb{V}\left[\mathcal{D}_{\text{R}}\right]=&\mathcal{L}_{\text{d}}^{2}\left[L_{1}\left(-\mu_{\text{d}}^{2}\right)-\frac{\pi}{4}L_{\frac{1}{2}}^{2}\left(-\mu_{\text{d}}^{2}\right)\right]+\mathcal{L}_{B}^{2}N_{2}{\mathcal{P}_{s}^{\zeta}}\left[\frac{1}{2}+\frac{Q}{4\pi}\sin\left(\frac{2\pi}{Q}\right)-{\mathcal{P}_{s}^{\zeta}}\frac{Q^{2}}{16}\sin^{2}\left(\frac{\pi}{Q}\right)\right]\nonumber\\
&+\frac{\mathcal{L}_{B}^{2}N_{2}(1-{\mathcal{P}_{s}^{\zeta}})}{4},
\end{align}
\begin{align}
\mathbb{V}\left[\mathcal{D}_{\text{I}}\right]=\mathcal{L}_{B}^{2}N_{2}{\mathcal{P}_{s}^{\zeta}}\left[\frac{1}{2}-\frac{Q}{4\pi}\sin\left(\frac{2\pi}{Q}\right)\right]+\frac{\mathcal{L}_{B}^{2}N_{2}(1-{\mathcal{P}_{s}^{\zeta}})}{4},
\end{align}
\begin{IEEEproof}
Please refer to Appendix A. 
\end{IEEEproof}

On the other hand, since the CSI for the interference channels through the $N_{B}$ REs is not available, the phase shifts for the interference signals are assumed to follow {the} uniform distribution as $\theta_{i}\sim U(0,2\pi)$, which leads to the random reflection of the interference signals\cite{9440664,10047976}. Let $\mathcal{I}_{k}\triangleq\left|\mathcal{F}_{k}\right|^{2}$ with $\mathcal{F}_{k}=\mathcal{L}_{\text{D}}^{(k)}h_{\text{D}}^{(k)}+\mathcal{L}_{B}^{(k)}\sum_{i=1}^{N_{B}}h_{i}^{(k)}g_{i}e^{j\theta_{i}}+\mathcal{L}_{B}^{(k)}\sum_{\ell=1}^{N_{b}}\hat{h}_{\ell}^{(k)}\hat{g}_{\ell}$, and thus for a practically large number of REs, $\mathcal{F}_{k}$ converges to a complex Gaussian RV as $\mathcal{F}_{k}\sim\mathcal{CN}\left(0,\sigma_{\mathcal{F}_{k}}^{2}\right)$, where by using \cite{9440664} and applying the steps similar to those in Appendix A, $\sigma_{\mathcal{F}_{k}}^{2}=\mathbb{V}\left[\mathcal{F}_{k}\right]$ is obtained as
\begin{align}
\sigma_{\mathcal{F}_{k}}^{2}&=\mathbb{V}\left[\mathcal{L}_{\text{D}}^{(k)}h_{\text{D}}^{(k)}\right]+\mathbb{V}\left[\mathcal{L}_{B}^{(k)}\textstyle\sum_{i=1}^{N_{B}}h_{i}^{(k)}g_{i}e^{j\theta_{i}}\right]+\mathbb{V}\left[\mathcal{L}_{B}^{(k)}\textstyle\sum_{\ell=1}^{N_{b}}\hat{h}_{\ell}^{(k)}\hat{g}_{\ell}\right]=(\mathcal{L}_{\text{D}}^{(k)})^{2}+(\mathcal{L}_{B}^{(k)})^{2}N_{2}.\label{SIGFN}
\end{align}

Therefore, $\mathcal{I}_{k}\triangleq\left|\mathcal{F}_{k}\right|^{2}$ follows the exponential distribution, where its MGF is written as
\begin{align}
    \mathcal{M}_{\mathcal{I}_{k}}\left(\mathcal{S}\right)=\frac{1}{1+\frac{\mathcal{S}}{\gamma_{th}}P_{k}\sigma_{\mathcal{F}_{k}}^{2}}.\label{MGFI}
\end{align}

By substituting (\ref{MGFD}) and (\ref{MGFI}) into (\ref{MGFmain}), and then by substituting (\ref{MGFmain}) into (\ref{P0}) and (\ref{P1}), $\mathcal{P}_{0}$ and $\mathcal{P}_{1}$ are obtained. We then substitute them into (\ref{Pbr}) to find the BER in the $r$-th repetition as 
\begin{align}
     \mathcal{P}_{b}=&\frac{1}{2}+\frac{1}{2\Gamma(M)}\int_{0}^{\infty}\mathcal{S}^{\frac{M}{2}-1}J_{M}\left(2\sqrt{\mathcal{S}}\right)e^{-\frac{\mathcal{S}N_{0}}{\gamma_{th}}}\prod_{k=1}^{K}\left(1+\frac{\mathcal{S}}{\gamma_{th}}P_{k}\sigma_{\mathcal{F}_{k}}^{2}\right)^{-1}\nonumber\\
     &\times \left[\left(1+\frac{\sigma_{1}^{2}\mathcal{S}\sigma_{\mathcal{D}}^{2}}{\gamma_{th}\mu_{\mathcal{D}} }\right)^{-\frac{\mu_{\mathcal{D}}^{2}}{\sigma_{\mathcal{D}}^{2}}}-\left(1+\frac{\sigma_{0}^{2}\mathcal{S}\sigma_{\mathcal{D}}^{2}}{\gamma_{th}\mu_{\mathcal{D}} }\right)^{-\frac{\mu_{\mathcal{D}}^{2}}{\sigma_{\mathcal{D}}^{2}}}\right]\text{d}\mathcal{S}.\label{PbrF}
\end{align}

Finally, by substituting (\ref{PbrF}) into (\ref{ABERsys}), the BER of the system with $\mathcal{R}$ repetitions is obtained. Furthermore, by exploiting \cite[Eq. (25.4.45)]{abram} and applying some algebraic manipulations, (\ref{PbrF}) can be approximated as
\begin{align}
    \mathcal{P}_{b}\approx &\frac{1}{2}+\frac{1}{2\Gamma(M)}\left(\frac{\gamma_{th}}{N_{0}}\right)^{\frac{M}{2}}\sum_{l=1}^{L}\omega_{l}\Omega_{l}^{\frac{M}{2}-1}J_{M}\left(2\sqrt{\frac{\Omega_{l}\gamma_{th}}{N_{0}}}\right)\prod_{k=1}^{K}\left(1+\frac{\Omega_{l}P_{k}\sigma_{\mathcal{F}_{k}}^{2}}{N_{0}}\right)^{-1}\nonumber\\
    &\times \left[\left(1+\frac{\sigma_{1}^{2}\Omega_{l}\sigma_{\mathcal{D}}^{2}}{N_{0}\mu_{\mathcal{D}} }\right)^{-\frac{\mu_{\mathcal{D}}^{2}}{\sigma_{\mathcal{D}}^{2}}}-\left(1+\frac{\sigma_{0}^{2}\Omega_{l}\sigma_{\mathcal{D}}^{2}}{N_{0}\mu_{\mathcal{D}} }\right)^{-\frac{\mu_{\mathcal{D}}^{2}}{\sigma_{\mathcal{D}}^{2}}}\right],
\end{align}
where $\Omega_{l}$ and $\omega_{l}$ are respectively the sample points and weights factors of the Laguerre orthogonal polynomial, tabulated in \cite[TABLE (25.9)]{abram}.  
\vspace{-10pt}\subsection{Lower-Bound BER Analysis}
In order to obtain a further simplified expression for the BER, we derive lower-bound expressions for the unconditional probabilities given in (\ref{P0}) and (\ref{P1}). Thus, first using the Jensen's inequality for MGFs, $\mathbb{E}\left[e^{zx}\right]\geq e^{z\mathbb{E}[x]}$\cite{THz42}, the MGF given in (\ref{MGFmain}) can be lower-bounded as
\begin{align}
    \mathcal{M}_{\sigma_{x,\chi}^{2}}\left(\mathcal{S}\right) &\geq e^{-\frac{\mathcal{S}N_{0}}{\gamma_{th}}}e^{-\frac{\mathcal{S}}{\gamma_{th}}\sigma_{\chi}^{2}\mathbb{E}\left[\mathcal{D}\right]}e^{-\frac{\mathcal{S}}{\gamma_{th}}\sum_{k=1}^{K}P_{k}\mathbb{E}\left[\mathcal{I}_{k}\right]}\nonumber\\
    &\geq e^{-\frac{\mathcal{S}}{\gamma_{th}}\left(\sigma_{\chi}^{2}\mu_{\mathcal{D}}+\sum_{k=1}^{K}P_{k}  \sigma_{\mathcal{F}_{k}}^{2}+N_{0}\right)},\label{upMGF}
\end{align}
where $\mu_{\mathcal{D}}$ and $\sigma_{\mathcal{F}_{k}}^{2}$ are given in (\ref{MEAND}) and (\ref{SIGFN}), respectively. 

By substituting (\ref{upMGF}) into (\ref{P0}), a lower-bound on the probability, $\mathcal{P}_{0}$, can be expressed as 
\begin{align}
    \mathcal{P}_{0}\geq & 1-\frac{1}{\Gamma(M)}\int_{0}^{\infty}\mathcal{S}^{\frac{M}{2}-1}J_{M}\left(2\sqrt{\mathcal{S}}\right) e^{-\frac{\mathcal{S}}{\gamma_{th}}\left(\sigma_{0}^{2}\mu_{\mathcal{D}}+\sum_{k=1}^{K}P_{k}  \sigma_{\mathcal{F}_{k}}^{2}+N_{0}\right)}\text{d}\mathcal{S}.\label{up1}
\end{align}

By setting $\mathcal{Z}\triangleq\frac{\mathcal{S}}{\gamma_{th}}\left(\sigma_{0}^{2}\mu_{\mathcal{D}}+\sum_{k=1}^{K}P_{k}  \sigma_{\mathcal{F}_{k}}^{2}+N_{0}\right)$, (\ref{up1}) is equivalently written as
\vspace{-10pt}\begin{align}
    &\mathcal{P}_{0}\geq  1-\frac{1}{\Gamma(M)}\left(\frac{\gamma_{th}}{\sigma_{0}^{2}\mu_{\mathcal{D}}+\sum_{k=1}^{K}P_{k}  \sigma_{\mathcal{F}_{k}}^{2}+N_{0}}\right)^{\frac{M}{2}} \nonumber\\
    &\times\int_{0}^{\infty} \mathcal{Z}^{\frac{M}{2}-1} J_{M}\left(2\sqrt{\frac{\gamma_{th}}{\sigma_{0}^{2}\mu_{\mathcal{D}}+\sum_{k=1}^{K}P_{k}  \sigma_{\mathcal{F}_{k}}^{2}+N_{0}}\mathcal{Z}}\right) \text{d}\mathcal{Z}.\label{intup}
\end{align}

By applying \cite[Eq. (8.353.2)]{bk} and \cite[Eq. (8.356.3)]{bk}, the integral of (\ref{intup}) {can be} evaluated, and after some algebraic manipulations, {the lower-bound on $\mathcal{P}_{0}$} is obtained as 
\begin{align}
    \mathcal{P}_{0}\geq \frac{1}{\Gamma(M)} \Gamma\left(M,\frac{\gamma_{th}}{\sigma_{0}^{2}\mu_{\mathcal{D}}+\sum_{k=1}^{K}P_{k}  \sigma_{\mathcal{F}_{k}}^{2}+N_{0}}\right).\label{upp0}
\end{align}

By following the same steps used for (\ref{upp0}), a lower-bound expression for $\mathcal{P}_{1}$ is obtained as
\begin{align}
    \mathcal{P}_{1}\geq 1- \frac{1}{\Gamma(M)} \Gamma\left(M,\frac{\gamma_{th}}{\sigma_{1}^{2}\mu_{\mathcal{D}}+\sum_{k=1}^{K}P_{k}  \sigma_{\mathcal{F}_{k}}^{2}+N_{0}}\right).\label{upp1}
\end{align}

By substituting (\ref{upp0}) and (\ref{upp1}) into (\ref{Pbr}), a lower-bound on the BER in the $r$-th repetition is obtained as
\vspace{-10pt}\begin{align}
    \mathcal{P}_{b}\geq \frac{1}{2}+\frac{\Gamma\left(M,\frac{\gamma_{th}}{\sigma_{0}^{2}\mu_{\mathcal{D}}+\sum_{k=1}^{K}P_{k}  \sigma_{\mathcal{F}_{k}}^{2}+N_{0}}\right)}{2\Gamma(M)} -\frac{\Gamma\left(M,\frac{\gamma_{th}}{\sigma_{1}^{2}\mu_{\mathcal{D}}+\sum_{k=1}^{K}P_{k}  \sigma_{\mathcal{F}_{k}}^{2}+N_{0}}\right)}{2\Gamma(M)}. \label{lowfind}
\end{align}

Finally, by substituting (\ref{lowfind}) into (\ref{ABERsys}), the lower-bound BER with $\mathcal{R}$ repetitions is obtained. 
\vspace{-5pt}\subsection{Optimal Threshold Selection}
 {As explained in Sec. II, in the proposed system, bit-0 and bit-1 are encoded at the Tx using low-value and high-value resistors, respectively, which generate thermal noise signals with distinct energy levels. The Rx detects the transmitted bit by comparing the energy of the received signal to a decision threshold. If the threshold is poorly chosen, it may result in incorrect bit detection—such as interpreting bit-0 as bit-1 and vice versa—thereby significantly degrading the detection accuracy and overall system performance. Thus, proper selection of the threshold value, $\gamma_{th}$, is crucial for ensuring reliable bit detection and accurately evaluating the system’s effectiveness.} To determine the optimal threshold which minimizes the BER, we rewrite the BER using (\ref{Pbr}), (\ref{cb0}), and (\ref{cb1}) as
\begin{align}
    {\mathcal{P}_{b}=\frac{1}{2}+\frac{1}{2\Gamma(M)} \Gamma\left(M,\frac{\gamma_{th}}{\sigma_{x,0}^2}\right) - \frac{1}{2\Gamma(M)} \Gamma\left(M,\frac{\gamma_{th}}{\sigma_{x,1}^2}\right).}\label{BERTHRESHH}
\end{align}

{From (\ref{BERTHRESHH}), considering the terms involving the upper incomplete gamma functions and the condition $\sigma_{x,0}^2 < \sigma_{x,1}^2$, it can be observed that for small values of $\gamma_{th}$, the BER decreases because the term $\Gamma\left(M, \frac{\gamma_{th}}{\sigma_{x,0}^2}\right)$ dominates and decays faster. However, for large values of $\gamma_{th}$, the BER increases as $\Gamma\left(M, \frac{\gamma_{th}}{\sigma_{x,1}^2}\right)$ decays more slowly due to $\sigma_{x,0}^2 < \sigma_{x,1}^2$. This transition ensures the existence of a single minimum\footnote{{This behavior of the BER in relation to the threshold values is also confirmed via numerical results presented in Sec. \ref{SECREsu}.}}. Therefore, it is established that $\mathcal{P}_b$ is a unimodal function of $\gamma_{th}$ \cite{convex}. Furthermore, based on the Fermat's theorem, any critical point of a differentiable function occurs where its first derivative is zero \cite{abram}.} Thus, to obtain the optimal threshold, we need to take the derivative of $\mathcal{P}_b$ with respect to $\gamma_{th}$, and set it to zero, i.e., $\frac{\partial \mathcal{P}_b}{\partial \gamma_{th}} = 0$.

As a result, using (\ref{BERTHRESHH}) and $\frac{\partial \Gamma\left(a,bx\right)}{\partial x}=-b(bx)^{a-1}e^{-bx}$\cite[Eq. (8.356.4)]{bk}, we have   
\begin{align}
    \frac{\partial \mathcal{P}_{b}}{\partial \gamma_{th}}&=\frac{\partial }{\partial \gamma_{th}}\left[\frac{1}{2\Gamma(M)}\Gamma\left(M,\frac{\gamma_{th}}{\sigma_{x,0}^{2}}\right)+\frac{1}{2}-\frac{1}{2\Gamma(M)}\Gamma\left(M,\frac{\gamma_{th}}{\sigma_{x,1}^{2}}\right)\right]\nonumber\\
    &=-\frac{1}{2\Gamma(M)\gamma_{th}}\left(\frac{\gamma_{th}}{\sigma_{x,0}^{2}}\right)^{M}e^{-\frac{\gamma_{th}}{\sigma_{x,0}^{2}}}+\frac{1}{2\Gamma(M)\gamma_{th}}\left(\frac{\gamma_{th}}{\sigma_{x,1}^{2}}\right)^{M}e^{-\frac{\gamma_{th}}{\sigma_{x,1}^{2}}}=0.\label{dPb1}
\end{align}

By applying some algebraic manipulations, (\ref{dPb1}) can be rewritten as
\begin{align}
    \ln\left[\left(\frac{1}{\sigma_{x,0}^{2}}\right)^{M}e^{-\frac{\gamma_{th}}{\sigma_{x,0}^{2}}}\right]&=\ln\left[\left(\frac{1}{\sigma_{x,1}^{2}}\right)^{M}e^{-\frac{\gamma_{th}}{\sigma_{x,1}^{2}}}\right]\nonumber\\
    -M\ln\sigma_{x,0}^{2}-\frac{\gamma_{th}}{\sigma_{x,0}^{2}}&=-M\ln\sigma_{x,1}^{2}-\frac{\gamma_{th}}{\sigma_{x,1}^{2}}\nonumber\\
    \gamma_{th}&=\frac{M\sigma_{x,0}^{2}\sigma_{x,1}^{2}}{\sigma_{x,1}^{2}-\sigma_{x,0}^{2}}\ln\left(\frac{\sigma_{x,1}^{2}}{\sigma_{x,0}^{2}}\right).\label{optn1}
\end{align}

Since (\ref{optn1}) involves $\sigma_{x,0}^{2}$ and $\sigma_{x,1}^{2}$ which are the {sums} of different RVs, and thus taking the expectation of (\ref{optn1}) for obtaining the average threshold, where the same RVs appear in both the numerator and denominator, is mathematically complex and cumbersome. Alternatively, using the lower-bound BER expression given in (\ref{lowfind}), and following the same steps used for (\ref{optn1}), an {approximated} expression for threshold selection can be obtained as 
\begin{align}
    \gamma_{th}\approx&\frac{M\left(\sigma_{0}^{2}\mu_{\mathcal{D}}+\sum_{k=1}^{K}P_{k}  \sigma_{\mathcal{F}_{k}}^{2}+N_{0}\right)\left(\mathcal{C}\sigma_{0}^{2}\mu_{\mathcal{D}}+\sum_{k=1}^{K}P_{k}  \sigma_{\mathcal{F}_{k}}^{2}+N_{0}\right)}{\sigma_{0}^{2}\mu_{\mathcal{D}}(\mathcal{C}-1)}\nonumber\\
    &\times\ln\left(\frac{\mathcal{C}\sigma_{0}^{2}\mu_{\mathcal{D}}+\sum_{k=1}^{K}P_{k}  \sigma_{\mathcal{F}_{k}}^{2}+N_{0}}{\sigma_{0}^{2}\mu_{\mathcal{D}}+\sum_{k=1}^{K}P_{k}  \sigma_{\mathcal{F}_{k}}^{2}+N_{0}}\right).\label{opttherap}
\end{align}
\begin{Remark}\label{REM2}From (\ref{opttherap}), it can be stated that the threshold selection depends on the number of REs capable of beamforming and blindly reflecting the signals, transmit power, and numbers of samples and interferers. 
\end{Remark}

\vspace{-20pt}\subsection{Mutual Information (MI)}\vspace{-5pt}
To gain further insights, we assess the MI between $s_{m,r}$ and $x_{m,r}$, which is expressed as
\vspace{-5pt}\begin{align}
    I\left(s_{m,r};x_{m,r}\right)=H\left(x_{m,r}\right)-H\left(x_{m,r}|s_{m,r}\right),\label{MI1}
\end{align}
where $H\left(x_{m,r}\right)$ and $H\left(x_{m,r}|s_{m,r}\right)$ are the marginal and conditional entropies, respectively. 

Since $s_{m,r}$ can be either bit-0 or bit-1 with equal probabilities, the received signal given in (\ref{rec1}) is a zero-mean complex Gaussian signal for both bit-0 and bit-1 with variances of $\sigma_{x,0}^{2}$ and $\sigma_{x,1}^{2}$, respectively, as given in (\ref{convar}). Therefore, the marginal entropy, $H\left(x_{m,r}\right)$, can be expressed as the average of the two marginal entropies for bit-0 and bit-1 as\cite{wu2014signal} 
\begin{align}
    H\left(x_{m,r}\right)&=\frac{1}{2}\Big[\mathbb{E}\left[\ln\left(\pi e \sigma_{x,0}^{2}\right)\right]+\mathbb{E}\left[\ln\left(\pi e \sigma_{x,1}^{2}\right)\right]\Big]\nonumber\\
    &=\frac{1}{2}\mathbb{E}\left[\ln\left(\pi e \left(\mathcal{D}\sigma_{0}^{2}+\displaystyle\sum_{k=1}^{K}P_{k}\mathcal{I}_{k}+N_{0}\right)\right)\right]\nonumber\\
    & \ \ \  +\frac{1}{2}\mathbb{E}\left[\ln\left(\pi e \left(\mathcal{D}\sigma_{1}^{2}+\displaystyle\sum_{k=1}^{K}P_{k}\mathcal{I}_{k}+N_{0}\right)\right)\right].\label{marg1}
\end{align}

The expectations in (\ref{marg1}) are with respect to the desired and interference channel gains along with the random numbers of REs, i.e., $\mathcal{D}$ and $\mathcal{I}_{k}$, where deriving their exact expectations are mathematically complex and {challenging}. However, using \cite[Eq. (17)]{NIPS2006_532b7cbe} and \cite[Sec. III-C]{10047976}, they can be accurately approximated via the Gaussian approximation based on the second-order Taylor expansion, i.e., $\mathbb{E}\left[\ln X\right]\approx \ln\mathbb{E}\left[X\right]-\frac{\mathbb{V}\left[X\right]}{2\mathbb{E}^{2}\left[X\right]}$. Therefore, the marginal entropy can be approximated as
\begin{align}
     H\left(x_{m,r}\right)\approx &\frac{1}{2}\ln\left(\pi e \mathbb{E}\left[\sigma_{x,0}^{2}\right]\right)-\frac{\mathbb{V}\left[\sigma_{x,0}^{2}\right]}{4\mathbb{E}^{2}\left[\sigma_{x,0}^{2}\right]} + \frac{1}{2}\ln\left(\pi e \mathbb{E}\left[\sigma_{x,1}^{2}\right]\right)-\frac{\mathbb{V}\left[\sigma_{x,1}^{2}\right]}{4\mathbb{E}^{2}\left[\sigma_{x,1}^{2}\right]},\label{marg2}
\end{align}
where $\sigma_{x,\chi}^{2}$, $\chi\in\{0,1\}$ is given in (\ref{convar}). 

In order to obtain (\ref{marg2}), {closed-form expressions for} the mean and variance, i.e., $\mathbb{E}\left[\sigma_{x,\chi}^{2}\right]$ and $\mathbb{V}\left[\sigma_{x,\chi}^{2}\right]$, {are required}. Therefore, using (\ref{convar}), we get
\begin{align} \mathbb{E}\left[\sigma_{x,\chi}^{2}\right]=\sigma_{\chi}^{2}\mathbb{E}\left[\mathcal{D}\right]+\textstyle\sum_{k=1}^{K}P_{k}\mathbb{E}\left[\mathcal{I}_{k}\right]+N_{0}=\sigma_{\chi}^{2}\mu_{\mathcal{D}}+\textstyle\sum_{k=1}^{K}P_{k}  \sigma_{\mathcal{F}_{k}}^{2}+N_{0}, \ \ \chi\in\{0,1\},
\end{align}
\begin{align}
\mathbb{V}\left[\sigma_{x,\chi}^{2}\right]=\mathbb{V}\left[\sigma_{\chi}^{2}\mathcal{D}\right]+\textstyle\sum_{k=1}^{K}\mathbb{V}\left[P_{k}\mathcal{I}_{k}\right]+\mathbb{V}\left[N_{0}\right]=\left(\sigma_{\chi}^{2}\right)^{2}\sigma_{\mathcal{D}}^{2}+\textstyle\sum_{k=1}^{K}P_{k}^{2}\left(\sigma_{\mathcal{F}_{k}}^{2}\right)^2, 
\end{align}
where $\mu_{\mathcal{D}}$, $\sigma_{\mathcal{D}}^{2}$ and $\sigma_{\mathcal{F}_{k}}^{2}$ are given in (\ref{MEAND}), (\ref{VARD}), and (\ref{SIGFN}), respectively. 

For computing the conditional entropy, the received signal is conditioned on $s_{m,r}$, and thus the uncertainty is due to the interference and noise components. As a result, similar to (\ref{marg2}), the conditional entropy can be expressed as 
\begin{align}
    H\left(x_{m,r}|s_{m,r}\right)&=\mathbb{E}\left[\ln\left[\pi e \left(\textstyle\sum_{k=1}^{K}P_{k}\mathcal{I}_{k}+N_{0}\right)\right]\right]\nonumber\\
    &\approx \ln\left[\pi e \ \mathbb{E}\left[\left(\textstyle\sum_{k=1}^{K}P_{k}\mathcal{I}_{k}+N_{0}\right)\right]\right]-\frac{\mathbb{V}\left[\pi e \left(\sum_{k=1}^{K}P_{k}\mathcal{I}_{k}+N_{0}\right)\right]}{2\mathbb{E}^{2}\left[\pi e \left(\sum_{k=1}^{K}P_{k}\mathcal{I}_{k}+N_{0}\right)\right]}\nonumber\\
     &\approx \ln\left[\pi e \left(\textstyle\sum_{k=1}^{K}P_{k}\sigma_{\mathcal{F}_{k}}^{2}+N_{0}\right)\right]-\frac{\sum_{k=1}^{K}P_{k}^{2}\left(\sigma_{\mathcal{F}_{k}}^{2}\right)^2}{2\left(\textstyle\sum_{k=1}^{K}P_{k}\sigma_{\mathcal{F}_{k}}^{2}+N_{0}\right)^{2}}.\label{condentF}
\end{align}

By substituting (\ref{marg2}) and (\ref{condentF}) into (\ref{MI1}), and after some algebraic manipulations, an {approximated} expression for the MI is obtained as 
\begin{align}
    I\left(s_{m,r};x_{m,r}\right)\approx & \frac{1}{2}\ln\left[\left(1+\frac{\sigma_{0}^{2}\mu_{\mathcal{D}}}{\mathcal{A}}\right)\left(1+\frac{\sigma_{1}^{2}\mu_{\mathcal{D}}}{\mathcal{A}}\right)\right]+\frac{\sum_{k=1}^{K}P_{k}^{2}\left(\sigma_{\mathcal{F}_{k}}^{2}\right)^2}{2\mathcal{A}^{2}}\nonumber\\
    &-\frac{\left(\sigma_{0}^{2}\right)^{2}\sigma_{\mathcal{D}}^{2}+\sum_{k=1}^{K}P_{k}^{2}\left(\sigma_{\mathcal{F}_{k}}^{2}\right)^2}{4\left(\sigma_{0}^{2}\mu_{\mathcal{D}}+\mathcal{A}\right)^{2}}-\frac{\left(\sigma_{1}^{2}\right)^{2}\sigma_{\mathcal{D}}^{2}+\sum_{k=1}^{K}P_{k}^{2}\left(\sigma_{\mathcal{F}_{k}}^{2}\right)^2}{4\left(\sigma_{1}^{2}\mu_{\mathcal{D}}+\mathcal{A}\right)^{2}},\label{MIAPROX}
\end{align}
where $\mathcal{A}=\sum_{k=1}^{K}P_{k}  \sigma_{\mathcal{F}_{k}}^{2}+N_{0}$.

Moreover, the MI over $M$ samples can be approximated as $I(S,X)\approx M I\left(s_{m,r};x_{m,r}\right)$, where the samples of the received signal are considered to be IID, as described in Section \ref{SECII}. 
\vspace{-5pt}\subsection{{Energy Efficiency (EE)}}\vspace{-5pt}
{In order to further evaluate the system performance, using (\ref{MIAPROX}), the EE of the system in terms of bits per joule (bit/J) can be written as \cite{10677372}}
\begin{align}
    {EE=\frac{I\left(s_{m,r};x_{m,r}\right)}{\ln2 \ P_{\text{Tot}}},}\label{EE1}
\end{align}
{where $I\left(s_{m,r};x_{m,r}\right)$ is the MI given in (\ref{MIAPROX}), and $P_{\text{Tot}}=P_{\text{Tx}}+P_{\text{RIS}}+P_{\text{Rx}}$ is the total power consumption, such that $P_{\text{Tx}}$, $P_{\text{RIS}}=N_{2}\mathcal{E}+\mathcal{E}_{con}$ as given in (\ref{con}), and $P_{\text{Rx}}$ are respectively the power consumptions at the Tx, RIS, and Rx\cite{10677372}.} 

{As discussed in Sec. \ref{SECII}, the Tx in the thermal noise modulation-based system does not require external energy sources; instead, it operates by switching between unbiased resistors through a digital logic controller, which consumes an extremely low amount of power, approximately 1 $\mu$W \cite{basar2024kirchhoff}. Moreover, the considered zero-energy RIS also does not require dedicated energy sources, as it harvests the required energy from existing RF interference signals. Thus, considering the power consumption of the Rx and the digital logic controller at the Tx, the total net power consumption becomes $P_{\text{Tot}}=P_{\text{Tx}}+P_{\text{Rx}}$. }
\subsection{Asymptotic Analysis}
As discussed in Section \ref{SECII}, the interference signals are exploited through EH at the RIS to provide the required power for the REs and beamforming. To gain deeper insights and investigate whether these signals are always beneficial, we analyze the asymptotic behavior of {the} system in the high-interference regime, i.e., $\{K,P_{k}\}\to \infty$. Thus, {using (\ref{ECSRmain}) and (\ref{ECNL1}),} the asymptotic ECSR {under both LEH and NLEH models,} ${\mathcal{P}_{s}^{\zeta}}^{\infty} = \lim_{K\to\infty}{\mathcal{P}_{s}^{\zeta}}$ {where $\zeta\in\{\text{L},\text{NL}\}$}, can be written as
\begin{align}
    {\mathcal{P}_{s}^{\zeta}}^{\infty}=\lim_{K\to\infty}\mathrm{P_{r}}\left(\textstyle\sum_{k=1}^{K} P_{k}(d_{1}^{(k)})^{-a}\left|\textstyle\sum_{n=1}^{N_{1}}\tilde{h}_{n}^{(k)}e^{j\tilde{\theta}_{n}}\right|^{2}\geq \mathcal{T}^{\zeta}\right)=1, \label{asyeh}
\end{align}
{where according to (\ref{ECSRmain}) and (\ref{ECNL1}), $\mathcal{T}^{\text{L}}=\frac{N_{2}\mathcal{E}+\mathcal{E}_{con}}{\eta}$ and $\mathcal{T}^{\text{NL}}=\frac{\mathcal{A}_{3}^{2}\left(N_{2}\mathcal{E}+\mathcal{E}_{con}\right)}{\mathcal{A}_{1}\mathcal{A}_{3}-\mathcal{A}_{3}\left(N_{2}\mathcal{E}+\mathcal{E}_{con}\right)-\mathcal{A}_{2}}$, respectively.} 

Furthermore, by exploiting (\ref{Pbr}), (\ref{cb0}) and (\ref{cb1}), the asymptotic BER, $\mathcal{P}_{b}^{\infty}= \lim_{K\to\infty}\mathcal{P}_{b}$, can be expressed as 
\begin{align}
    \mathcal{P}_{b}^{\infty}&=\frac{1}{2}+\lim_{K\to\infty} \frac{1}{2\Gamma(M)}\Gamma\left(M,\frac{\gamma_{th}}{\mathcal{D}\sigma_{0}^{2}+\sum_{k=1}^{K}P_{k}\mathcal{I}_{k}+N_{0}}\right)\nonumber\\
    &\ \ \ -\lim_{K\to\infty} \frac{1}{2\Gamma(M)}\Gamma\left(M,\frac{\gamma_{th}}{\mathcal{D}\sigma_{1}^{2}+\sum_{k=1}^{K}P_{k}\mathcal{I}_{k}+N_{0}}\right)\nonumber\\
    &=\frac{1}{2}+\frac{1}{2\Gamma(M)}\Gamma\left(M,0\right)-\frac{1}{2\Gamma(M)}\Gamma\left(M,0\right)=\frac{1}{2}.\label{ASBER}
\end{align}

Using (\ref{MI1}), (\ref{marg1}), (\ref{condentF}) and $\ln(1)=0$, the asymptotic MI, $I^{\infty}\left(s_{m,r};x_{m,r}\right)=\lim_{K\to\infty}I\left(s_{m,r};x_{m,r}\right)$ is obtained as
\begin{align}
    I^{\infty}\left(s_{m,r};x_{m,r}\right)&=\lim_{K\to\infty} \left[H\left(x_{m,r}\right)-H\left(x_{m,r}|s_{m,r}\right)\right]\nonumber\\
    &=\lim_{K\to\infty} \mathbb{E}\left[\frac{1}{2}\ln\left(\pi^{2} e^{2} \left(\mathcal{D}\sigma_{0}^{2}+\textstyle\sum_{k=1}^{K}P_{k}\mathcal{I}_{k}+N_{0}\right)\left(\mathcal{D}\sigma_{1}^{2}+\textstyle\sum_{k=1}^{K}P_{k}\mathcal{I}_{k}+N_{0}\right)\right)\right.\nonumber\\
    & \left. \ \ \ \ -\ln\left(\pi e \left(\textstyle\sum_{k=1}^{K}P_{k}\mathcal{I}_{k}+N_{0}\right)\right)\right]\nonumber\\
    &=\frac{1}{2}\lim_{K\to\infty}\mathbb{E}\left[\ln\left(1+\frac{\mathcal{D}\sigma_{0}^{2}}{\sum_{k=1}^{K}P_{k}\mathcal{I}_{k}+N_{0}}\right)+\ln\left(1+\frac{\mathcal{D}\sigma_{1}^{2}}{\sum_{k=1}^{K}P_{k}\mathcal{I}_{k}+N_{0}}\right)\right]=0.\label{ASMI}
\end{align}

{Similarly, using (\ref{EE1}) and (\ref{ASMI}), the asymptotic EE, $EE^{\infty} = \lim_{K\to\infty}EE$, is obtained as}
\begin{align}
    {EE^{\infty} = \lim_{K\to\infty}\frac{I\left(s_{m,r};x_{m,r}\right)}{\ln2 \ (P_{\text{Tx}}+P_{\text{Rx}})}=\frac{I^{\infty}\left(s_{m,r};x_{m,r}\right)}{\ln2 \ (P_{\text{Tx}}+P_{\text{Rx}})}=0.}\label{ASEE}
\end{align}

It is noteworthy that the asymptotic expressions remain valid when the interference power approaches infinity, i.e., $P_{k} \to \infty$.
\begin{Remark}\label{REM3}According to (\ref{ASBER}), (\ref{ASMI}), and (\ref{ASEE}), the communication performance for {the case with} a very large number of interferers and high interference power is severely limited at the Rx. 
\end{Remark}
\section{Results and Discussion}\label{SECREsu}
\vspace{-10pt}We present numerical and simulation results to validate the accuracy of the derived expressions and assess the system performance under various conditions, using the following parameters: $\{d,d_{1},d_{2}\}=\{8,3,6\}$ m, $\{d_{1}^{(1)},d_{1}^{(2)},d_{1}^{(3)},d_{1}^{(4)}\}=\{12,14,18,20\}$ m, $\{d_{\text{D}}^{(1)},d_{\text{D}}^{(2)},d_{\text{D}}^{(3)},d_{\text{D}}^{(4)}\}=\{18,20,22,25\}$ m, $a=2$, $\mu_{\text{d}}=1$, $\mathcal{C}=15$, $\eta=0.9$, $\{\mathcal{E},\mathcal{E}_{con}\}=\{1,50\}$ mW \cite{10348506,10677372}, $b=2$ bits, $N_{0}=-100$ dBm, and $\gamma_{th}$ is set as (\ref{opttherap}), unless otherwise stated. The fading channels and the RIS phase shifts in the simulation results are set as shown in TABLE \ref{TAB2}.

{Figs. \ref{FigECSR}, \ref{Fig3} and \ref{Fig4}} respectively show the {impacts} of REs allocation on the ECSR, BER, and MI under both ideal and quantized RIS phase shifts with $N=200$, $K=\{2,4\}$, $\sigma_{0}^{2}=-10$ dBm, $P_{k}=20$ dBm, $M=15$, and $\mathcal{R}=5$. The close match between the theoretical and simulation results confirms the accuracy of the expressions given in (\ref{pofeh}), (\ref{PbrF}), and (\ref{MIAPROX}). As shown in Fig. \ref{FigECSR}, for a given number of REs allocated to EH, an increase in the number of interferers increases the ECSR, thereby enhancing the amount of harvested energy required for beamforming. For example, for $N_{1}=125$, an ECSR of ${\mathcal{P}_{s}^{\zeta}}=0.57$ is achieved when $K=2$, and ${\mathcal{P}_{s}^{\zeta}}\approx 1$ when $K=4$, which {confirms} Remark \ref{rem1}. Furthermore, increasing $N_{1}$ raises the ECSR, reaching ${\mathcal{P}_{s}^{\zeta}}=1$ at approximately $N_{1}\geq 125$ for $K=4$, as more REs for EH enable sufficient energy harvesting. In contrast, allocating a large number of REs to signal reflection does not always guarantee the optimal BER and MI. For example, with $K=4$ and $N_2 =N-N_1= 125$, the BER and MI are about $10^{-1}$ and $0.7$ nats, respectively, as shown in Figs. \ref{Fig3} and \ref{Fig4}. This results from fewer REs allocated to EH ($N_1 = 75$), leading to an ECSR of roughly $0.17$—insufficient energy for beamforming at the RIS. Consequently, on average, only $\mathbb{E}[N_B] = {\mathcal{P}_s^\zeta} N_2 \approx 21$ REs support beamforming, while the remaining $\mathbb{E}[N_b] = (1 - {\mathcal{P}_s^\zeta}) N_2 \approx 104$ REs perform blind reflection.

It is observed that increasing $N_{1}$ improves the BER and MI until the ECSR reaches its maximum. Specifically, for $K=4$, at approximately $\{N_{1}, N_{2}\} = \{125, 75\}$—resulting in a maximized ECSR of 1—the system achieves the minimum BER and maximum MI. This outcome {confirms that the use of Algorithm \ref{ALG1} yields the optimal allocation of RIS resources for EH and communication}. However, increasing $N_1$ beyond 125 deteriorates both the BER and MI. Although this maximizes the ECSR—ensuring sufficient energy for RIS beamforming—it reduces the number of REs for reflection ($N_2$), thereby weakening the received signal due to improper resource allocation. Moreover, comparisons of the BER and MI achieved under ideal and quantized phases with $b=2$ bits show the {severe} impacts of quantization errors, which are common in practical RIS setups equipped with cost-effective discrete phase shifters\cite{10328394}.
\begin{figure*}[t]
  \centering
  \begin{subfigure}[b]{0.32\textwidth}
    \centering
    \includegraphics[width=\linewidth]{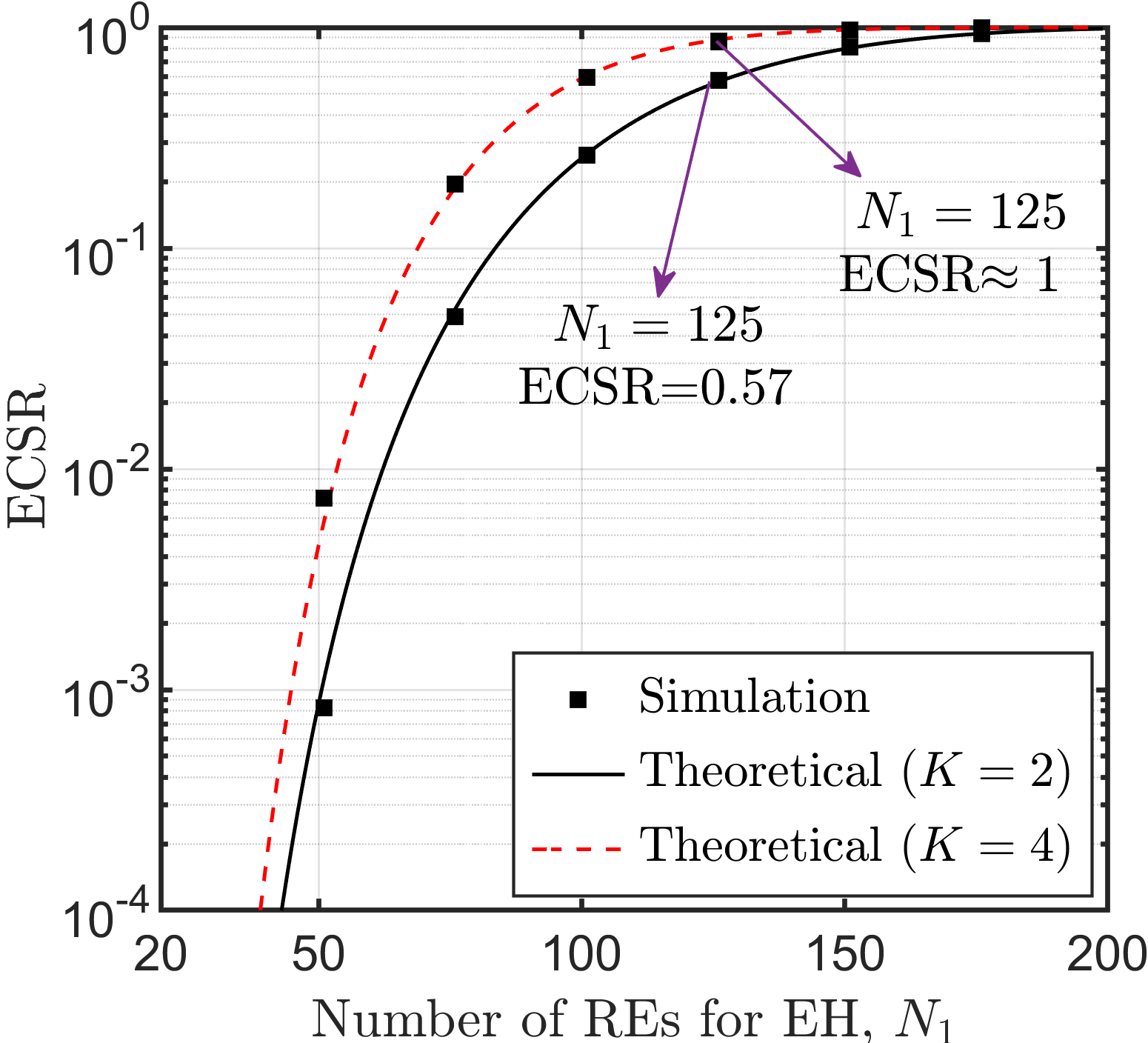}
    \vspace{-25pt}\caption{}
    \label{FigECSR}
  \end{subfigure}
  \hfill
  \begin{subfigure}[b]{0.32\textwidth}
    \centering
    \includegraphics[width=\linewidth]{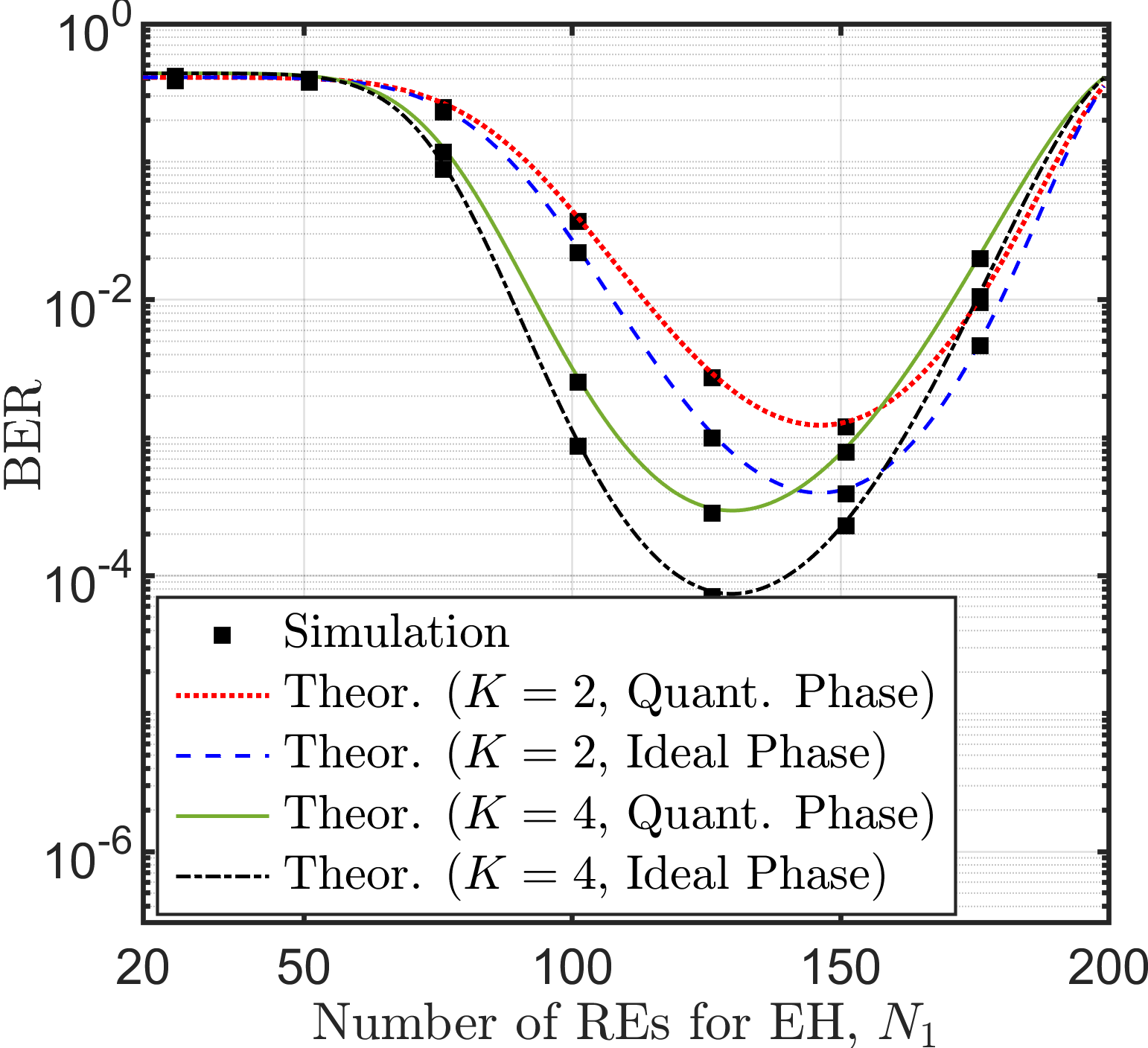}
    \vspace{-25pt}\caption{}
    \label{Fig3}
  \end{subfigure}
  \hfill
  \begin{subfigure}[b]{0.32\textwidth}
    \centering
    \includegraphics[width=\linewidth]{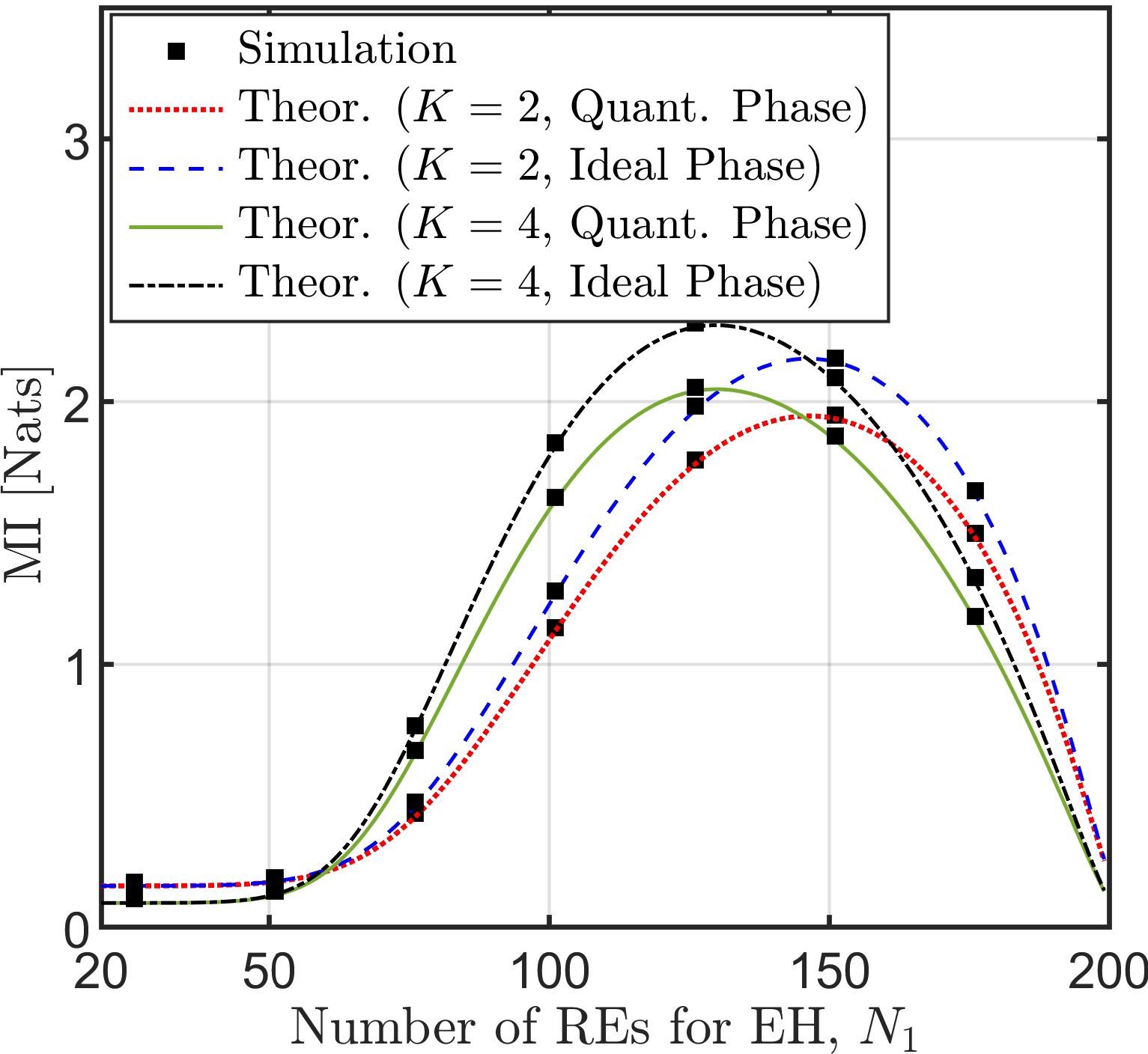}
    \vspace{-25pt}\caption{}
    \label{Fig4}
  \end{subfigure}
 \vspace{-15pt}\caption{%
     \ Effects of REs for EH on the system performance: (a) ECSR, (b) BER, and (c) MI.
  }
  \label{FigAll}\vspace{-30pt}
\end{figure*}

Fig. \ref{Fig5} shows the effects of repetition coding and SNR on the BER, where the SNR and interference-to-noise ratio are set as $\gamma_{0}=\sigma_{0}^{2}/N_{0}$ and $\tilde{\gamma_{k}}=P_{k}/N_{0}=5$ dB, respectively. {As expected, the results show that increasing $\gamma_{0}$ significantly reduces the BER, which can be theoretically achieved by increasing the noise-generating resistors, $R_{L}$ and $R_{H}$, at the Tx. However, in practical noise modulation systems, achieving a high transmit SNR is often constrained by energy and hardware limitations. A more practical solution is to apply the repetition coding, which introduces time diversity. As illustrated in the figure, at $\gamma_{0} = -5$ dB, the BER drops from approximately 0.046 without coding to 0.006 with $\mathcal{R} = 3$ repetitions, and further to $8.8\times10^{-4}$ with $\mathcal{R} = 5$. This improvement results from transmitting the same symbol multiple times using repetition coding, allowing the Rx to average multiple energy measurements rather than relying on a single observation, which yields a more accurate estimate of the transmitted noise power level and reduces uncertainty in the detection process—without increasing the Tx power.}

To find the optimal threshold that minimizes the BER, using (\ref{ABERsys}), (\ref{PbrF}), and (\ref{lowfind}), the accurate and lower-bound BER results for a range of selected thresholds, $\gamma_{th}$, under different numbers of samples and REs, are {illustrated} in Fig. \ref{Fig6}. Moreover, to verify the accuracy of the {approximated} expression for properly selecting the threshold given in (\ref{opttherap}), the BER achieved using this approximation is also depicted for comparison purposes, confirming its accuracy across different numbers of samples and REs. It is observed that the optimal threshold varies with the variations in the numbers of samples and REs, which is consistent with Remark \ref{REM2}. For instance, the system achieves the minimum BER with $\{M, N_{2}\} = \{4, 20\}$, $\{M, N_{2}\} = \{8, 20\}$, and $\{M, N_{2}\} = \{8, 40\}$ at the threshold values of approximately $9.5$ dB, $12.5$ dB, and $15$ dB, respectively. Furthermore, Fig. \ref{Fig6} shows that increasing the number of samples improves the BER performance. For instance, as shown in the figure, within a reasonable threshold range and with $N_{2}=20$, the BER improves as $M$ increases from 4 to 8. This is because a greater number of samples provides a more reliable estimate of the total energy in the received signal, enhancing the detection performance. In addition, increasing the number of RIS REs used for signal reflection enhances the system performance. For example, {for the same $M=8$, the BER achieved with $N_{2} = 20$ is significantly improved} when $N_{2}$ is increased to 40, as the RIS reflects signals {with} more REs, {introducing more degrees of freedom to improve the receive signal strength.}

{Fig.~\ref{Figx} depicts the EE performance of the proposed integration under LEH and NLEH models, compared to a conventional RIS-assisted system without EH capability. In this setup, we set $P_{\text{Rx}} = 195$~mW, reflecting the typical power consumption of an SDR-based Rx used in passive wireless communication systems~\cite{9982299}. We also set $P_{\text{Tx}} = 1\,\mu\text{W}$ consumed by the digital logic controller of the proposed integration \cite{basar2024kirchhoff}, and $P_{\text{Tx}} = (\sigma_0^2 + \sigma_1^2)/2$ for the conventional system due to its reliance on an active Tx. Fig.~\ref{Figx} demonstrates that the proposed system outperforms the conventional system under both LEH and NLEH models. Although the EE under the NLEH model is slightly lower than that of the LEH model, it still yields substantial gains over the conventional system and offers a more practical implementation by accounting for the non-linear behavior of practical EH circuitry.}
\begin{figure*}[t]
\begin{multicols}{3}
    \includegraphics[width=0.99\linewidth]{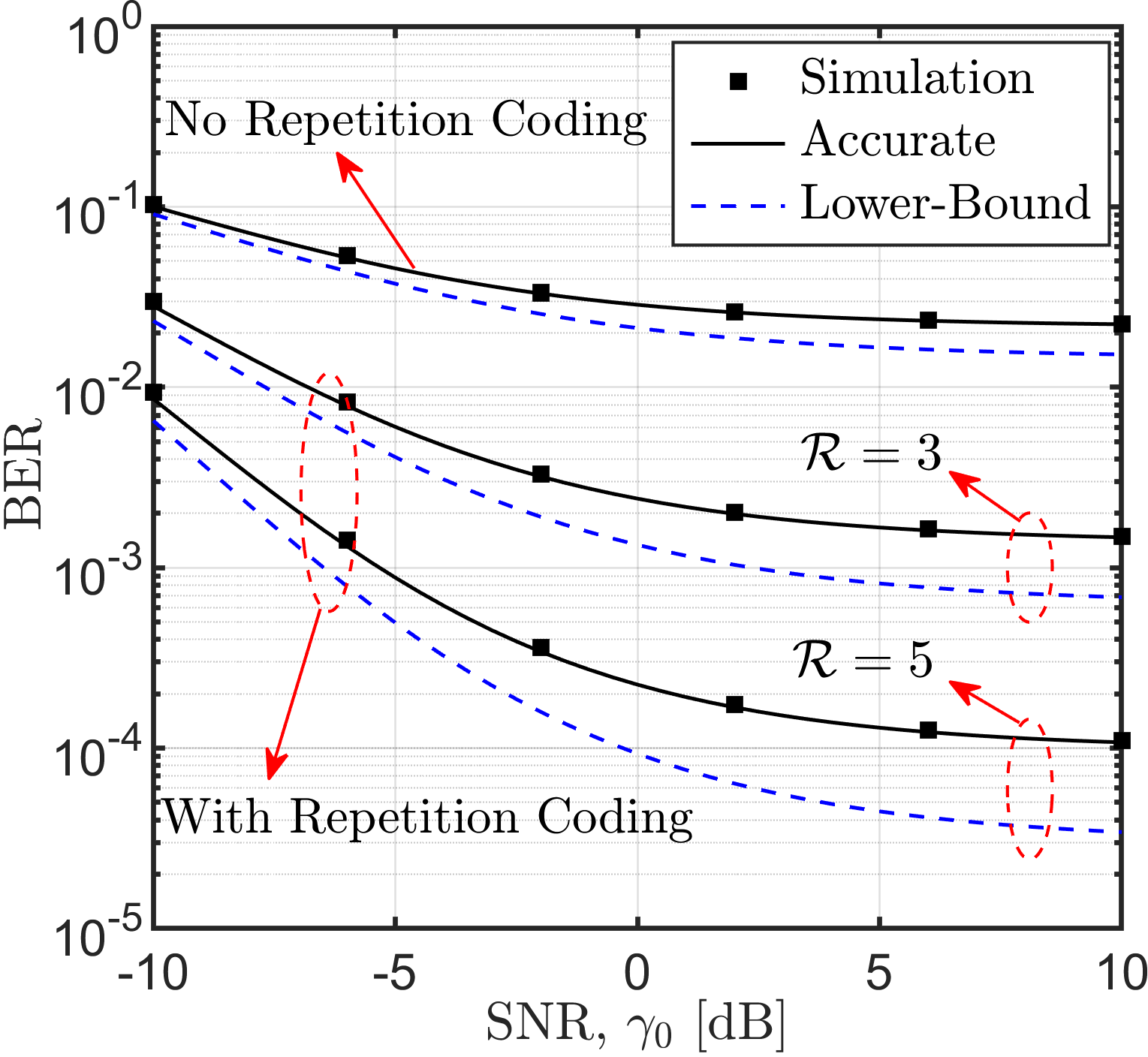}\par \vspace{-10pt}\caption{\ Effects of repetition coding and SNR on the BER for $N_{2}=50$, ${\mathcal{P}_{s}^{\text{L}}}=0.9$, $K=3$, and $M=3$.}\label{Fig5}
   \includegraphics[width=0.99\linewidth]{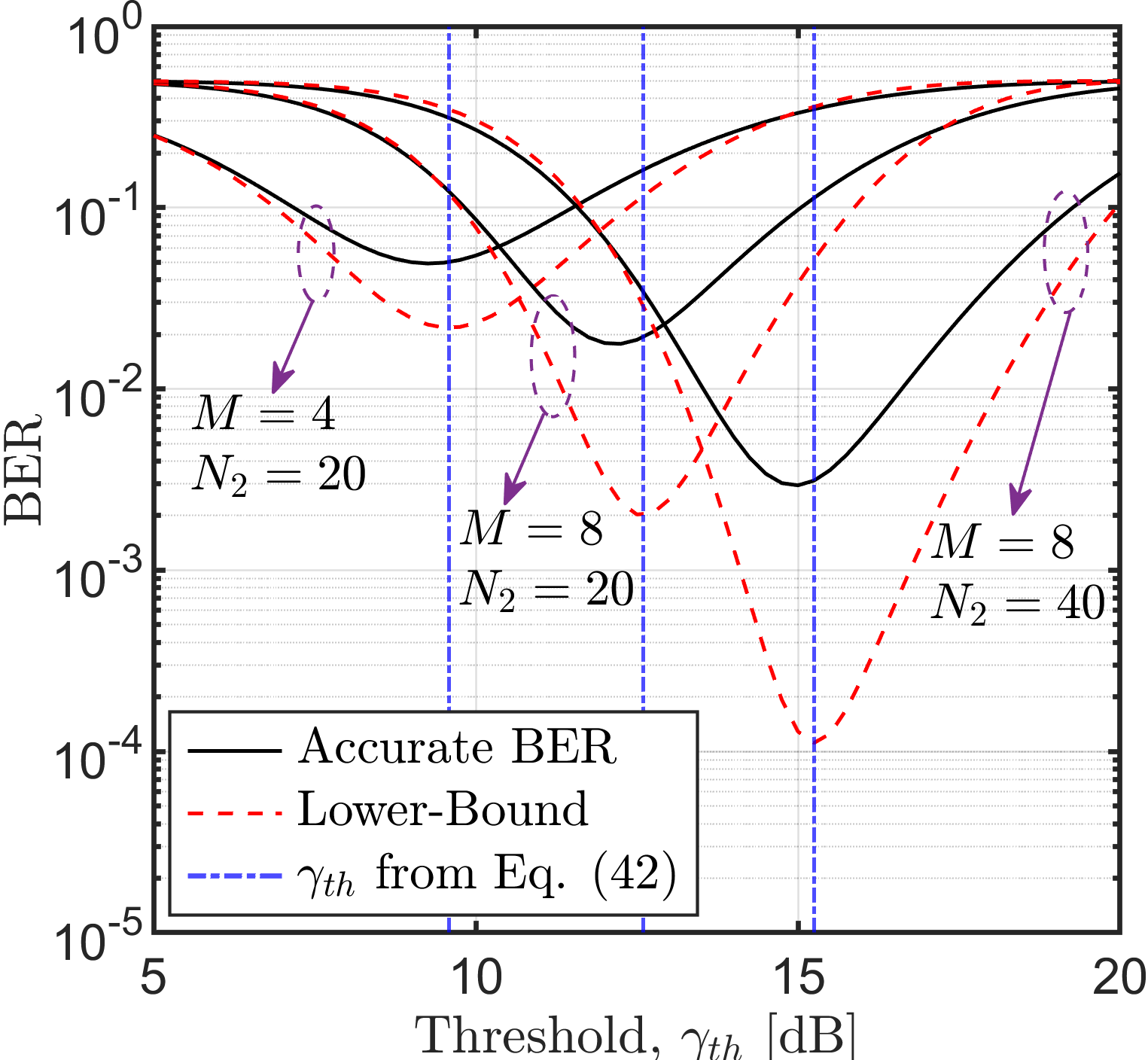}\par \vspace{-10pt}\caption{\ Effects of threshold values on the BER for $K=\mathcal{R}=3$, ${\mathcal{P}_{s}^{\text{L}}}=0.5$, and $\{\gamma_{0},\tilde{\gamma}_{k}\}=\{1,5\}$ dB.}\label{Fig6}
   \includegraphics[width=0.99\linewidth]{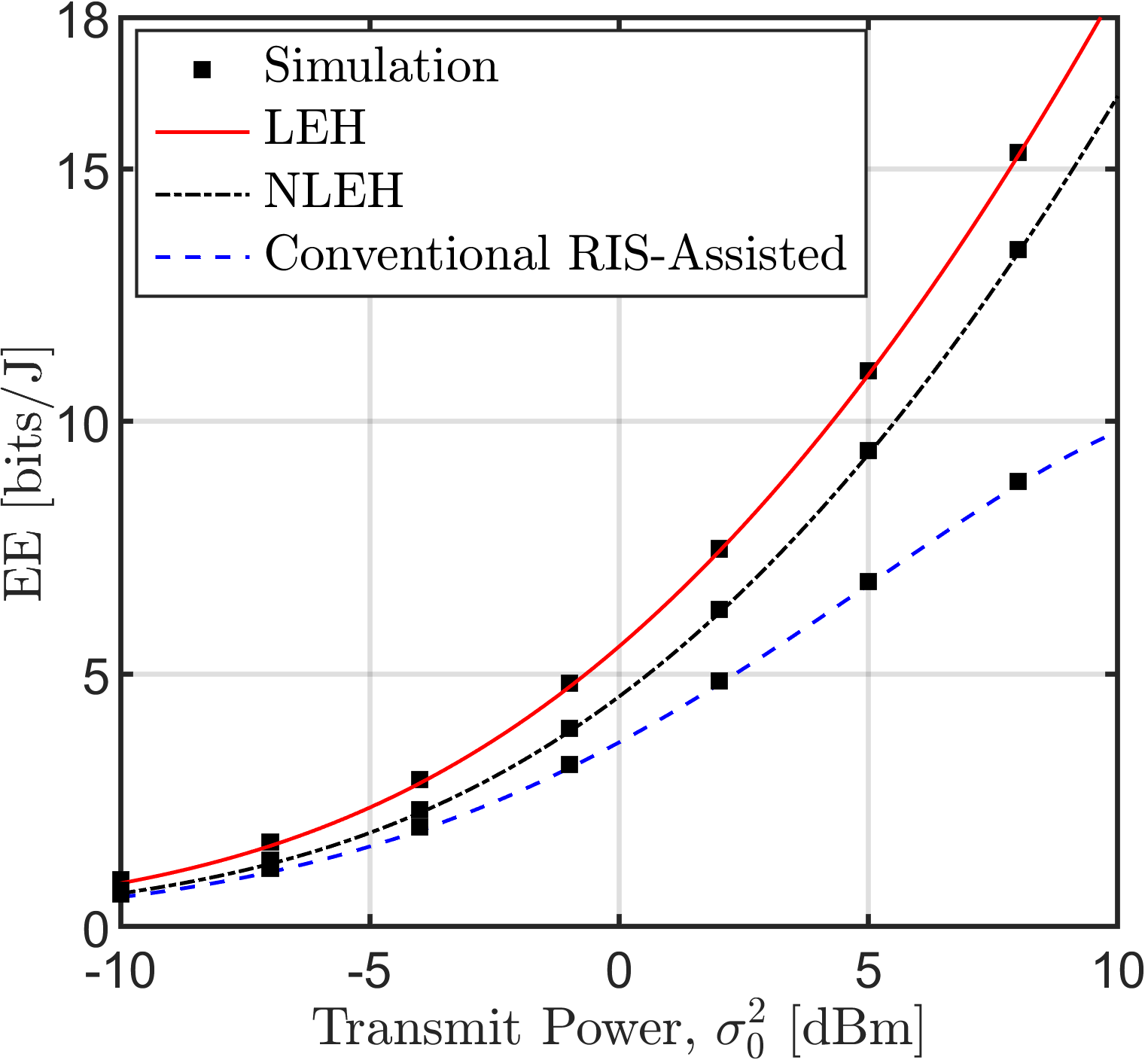}\par \vspace{-10pt}\caption{\ {EE Comparison of EH models with conventional RIS for $N_{2}=50$, $\eta=0.5$, and $K=3$.}}\label{Figx}
\end{multicols}\vspace{-40pt}
\end{figure*}

{Figs. \ref{FigECSRLNL}, \ref{Fig3LNL}, and \ref{Fig4LNL} present a comparative evaluation of the LEH and NLEH, highlighting their influence on the REs allocation and performance. The results are obtained by setting $N = 300$, $K = 4$, $b = 2$ bits, and $\eta = \{0.5, 0.75, 1\}$, while the remaining parameters are the same as those used in Figs.~\ref{FigECSR}, \ref{Fig3}, and~\ref{Fig4}. The LEH model assumes a linear relationship between the incident RF power and harvested energy, with performance scaling proportionally with $\eta$. In contrast, the NLEH model, which more accurately captures practical EH circuits, accounts for saturation effects and does not include $\eta$ as a tunable parameter. As a result, the two models lead to notably different RE allocation requirements to satisfy the energy demands for RIS beamforming.} 
\begin{figure*}[t]
  \centering
  \begin{subfigure}[b]{0.32\textwidth}
    \centering
    \includegraphics[width=\linewidth]{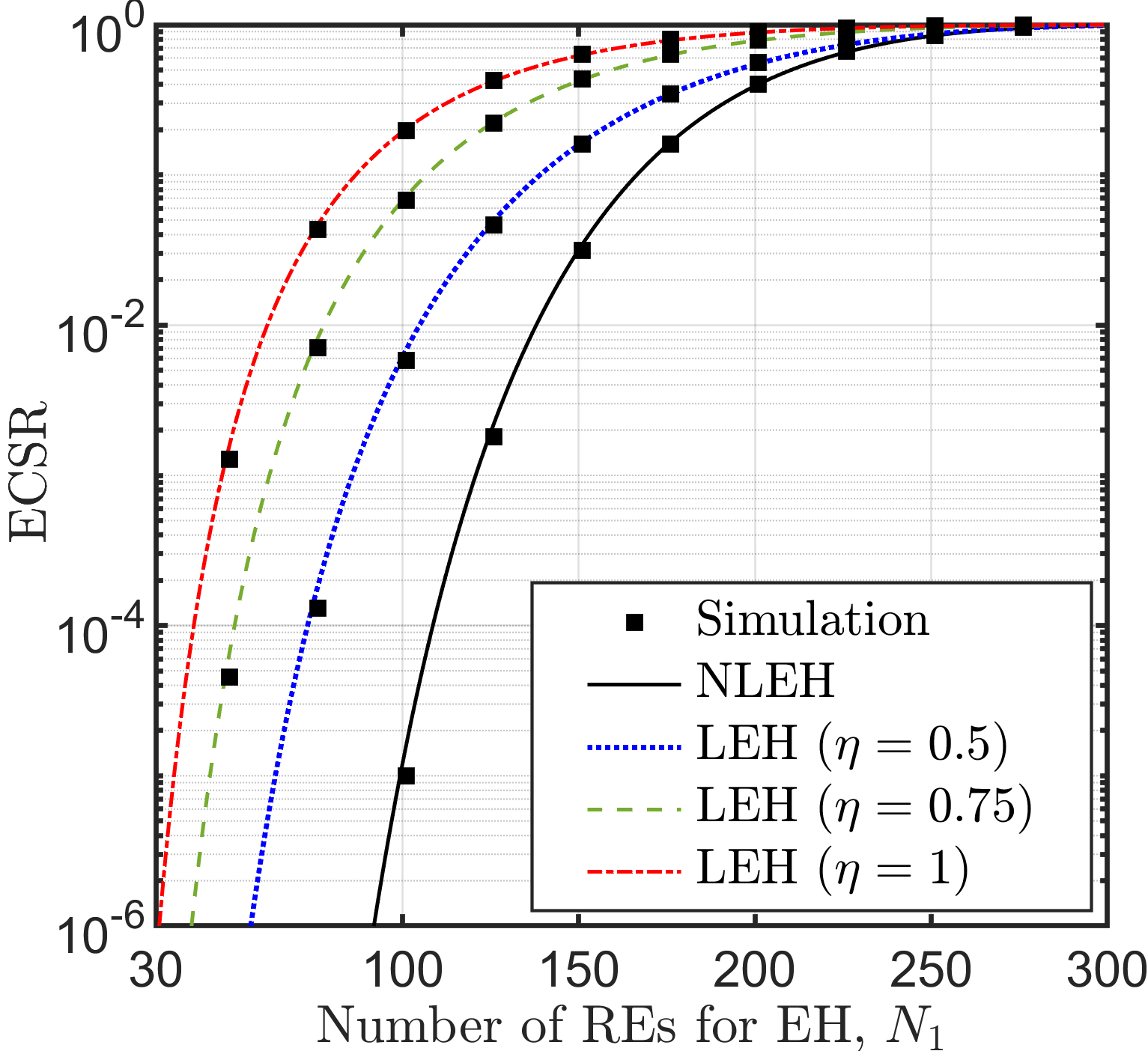}
   \vspace{-25pt} \caption{}
    \label{FigECSRLNL}
  \end{subfigure}
  \hfill
  \begin{subfigure}[b]{0.32\textwidth}
    \centering
    \includegraphics[width=\linewidth]{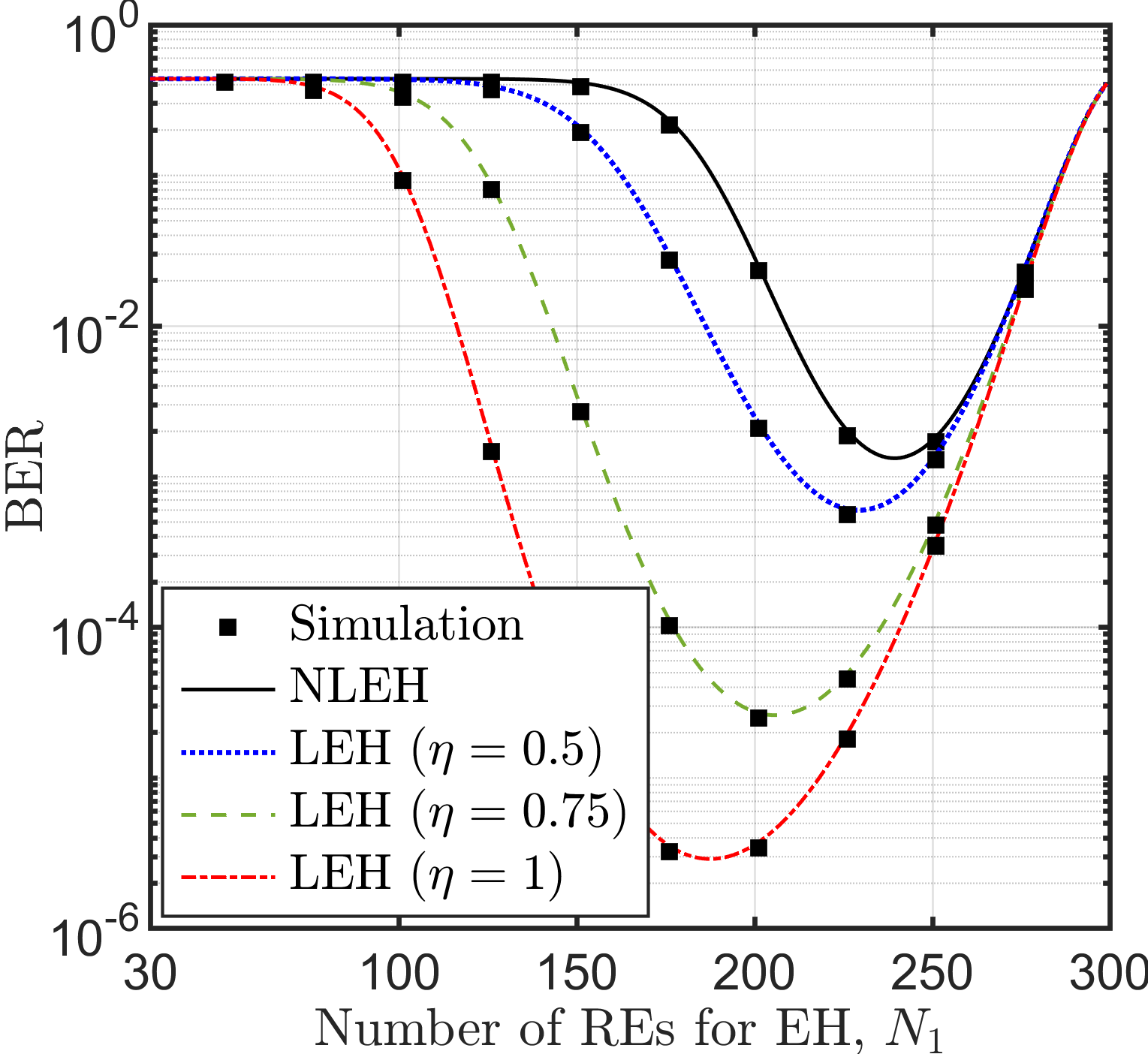}
    \vspace{-25pt}\caption{}
    \label{Fig3LNL}
  \end{subfigure}
  \hfill
  \begin{subfigure}[b]{0.32\textwidth}
    \centering
    \includegraphics[width=\linewidth]{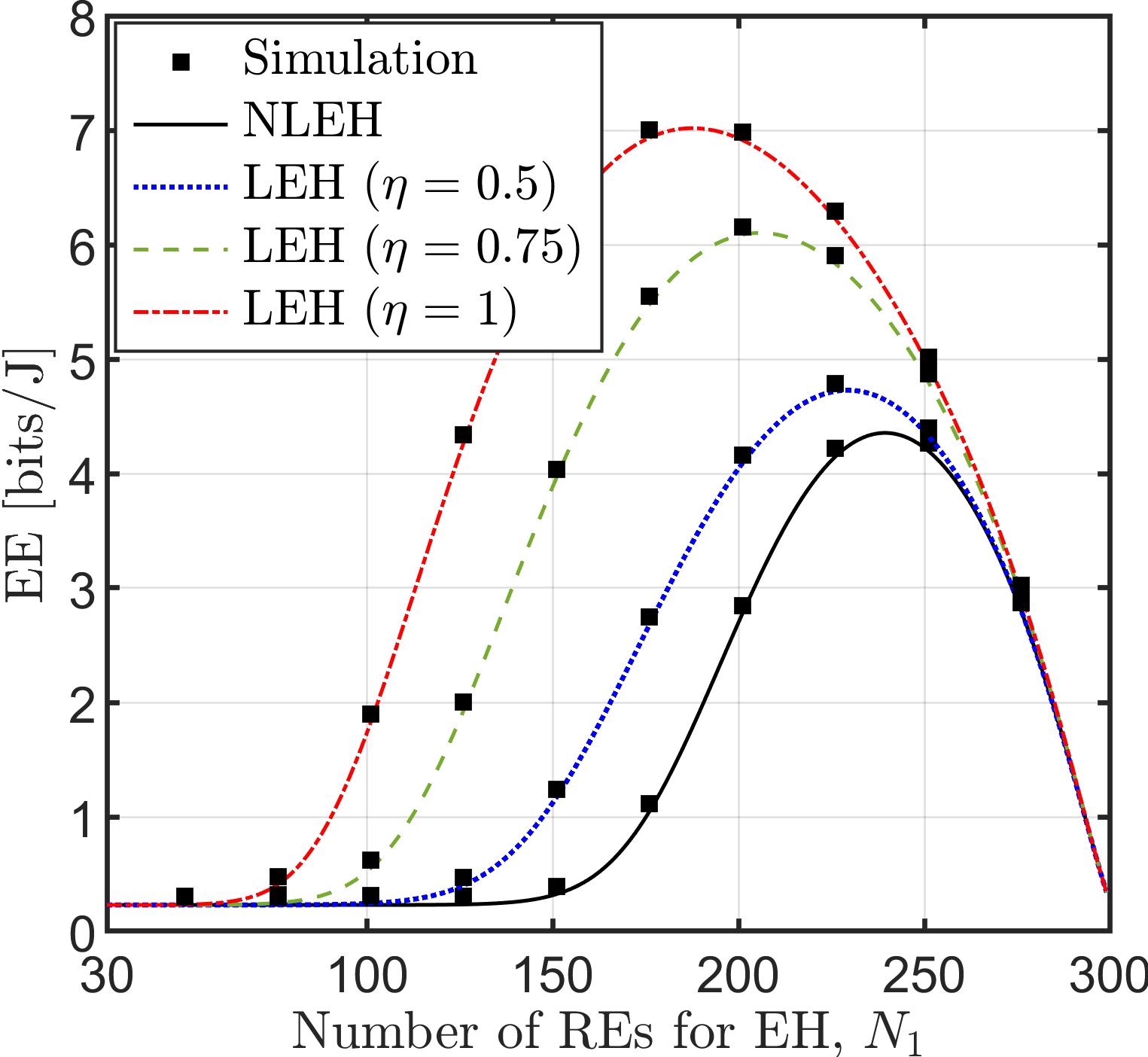}
   \vspace{-25pt} \caption{}
    \label{Fig4LNL}
  \end{subfigure}
\vspace{-15pt}\caption{\ {Effects of LEH and NLEH models on the system performance: (a) ECSR, (b) BER, and (c) EE.}}
  \label{FigAllLNL}\vspace{-30pt}
\end{figure*}

{As illustrated in Fig.~\ref{FigECSRLNL}, the ECSR under LEH increases rapidly with $N_1$, reaching its maximum with a relatively small number of REs allocated to EH; for example, at $N_1 \geq 200$ for $\eta = 0.75$. This enables more REs to be retained for signal reflection ($N_2 = 100$), resulting in improved signal quality via RIS beamforming. In contrast, the ECSR under NLEH increases more gradually and reaches its maximum at significantly larger values of $N_1$ (e.g., $N_1 \geq 240$), reflecting the limited harvesting capability of realistic EH circuits. These differences in harvested energy availability directly influence the BER and EE performance, as shown in Figs.~\ref{Fig3LNL} and~\ref{Fig4LNL}. Specifically, the optimal $N_1$ that minimizes BER and maximizes EE shifts toward higher values under NLEH, as more REs must be allocated to EH to meet the energy constraint required for RIS beamforming. For instance, the minimum BER under LEH with $\eta = 0.75$ occurs at $N_1 \approx 200$, where $\text{BER} \approx 2.4\times10^{-5}$, whereas under NLEH, the minimum BER of approximately $10^{-3}$ is achieved at $N_1 = 240$. Moreover, the maximum achievable EE under NLEH is consistently lower, due to reduced energy conversion efficiency and the consequent reduction in available REs for beamforming. For example, the maximum EE under LEH with $\eta = 0.75$ is attained at $N_1 = 200$, where $\mathcal{P}_{s}^{\text{L}}=1$ and $\text{EE} \approx 6$~bits/J, while under NLEH, the EE peaks later at $N_1 = 240$, reaching only $\text{EE} \approx 4.4$~bits/J. This performance degradation arises from the need to allocate a larger fraction of REs to EH under NLEH, leaving fewer REs available for signal reflection. These findings underscore the critical role of the EH model in determining the optimal trade-off between EH and communication performance. While LEH offers a useful performance bound, it does not account for practical nonlinearity constraints. Conversely, the NLEH model, provides realistic insights essential for the design of energy-aware and reliable RIS-assisted networks.}

{Fig.~\ref{FIGComplex} compares the computational complexity of the proposed binary search algorithm with exhaustive search and random search methods. The proposed algorithm requires \( \mathcal{O}(\log_2 N) \) complexity by successively halving the search space over \( N_1 \in \{1, 2, \ldots, N-1\} \) to find the optimal REs allocation. While classical binary search requires at most \( \lceil \log_2(N) \rceil \) evaluations, our implementation performs two additional evaluations due to problem-specific requirements: at line 8, where the algorithm continues after finding a feasible \( N_1 \) to seek the smallest such value, and at line 4, to ensure the final candidate is checked. Thus, the total number of ECSR computations is \( \lceil \log_2(N) \rceil + 2 \). For instance, both \( N = 300 \) and \( N = 500 \) yield 11 evaluations, matching \( \lceil \log_2(300) \rceil = \lceil \log_2(500) \rceil = 9 \), plus two extra. In contrast, an exhaustive search evaluates all \( N - 1 \) splits until a feasible point is found or exhausted, resulting in \( \mathcal{O}(N) \) complexity. The random search method randomly permutes the \( N - 1 \) possible splits and tests them sequentially until the first valid \( N_1 \) is found. Assuming the optimal point is equally likely to appear in any position, the expected number of evaluations is approximately \( \mathbb{E}[X] = (N - 1 + 1)/2 = N/2 \). These results highlight the substantial computational advantage of the proposed algorithm.}
\begin{figure*}[t]
\begin{multicols}{3}
    \includegraphics[width=0.99\linewidth]{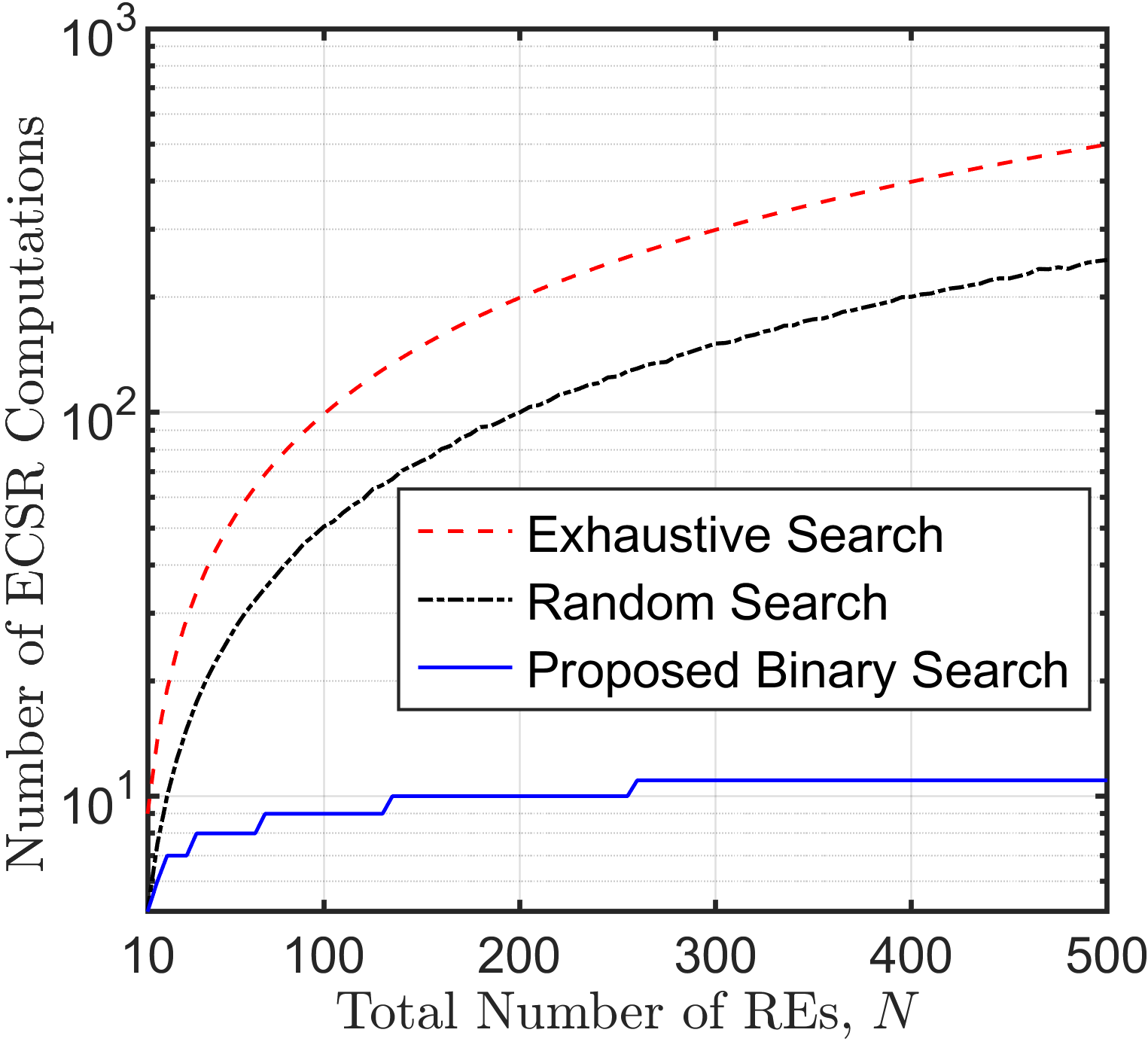}\par \vspace{-10pt}\caption{\ {Computation complexity of search algorithms.}}\label{FIGComplex}
    \includegraphics[width=0.99\linewidth]{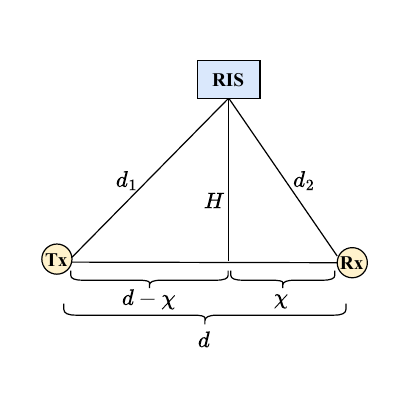}\par \vspace{-19pt}\caption{\ Simulation setup for RIS-assisted system.}\label{Fig7}
   \includegraphics[width=0.98\linewidth]{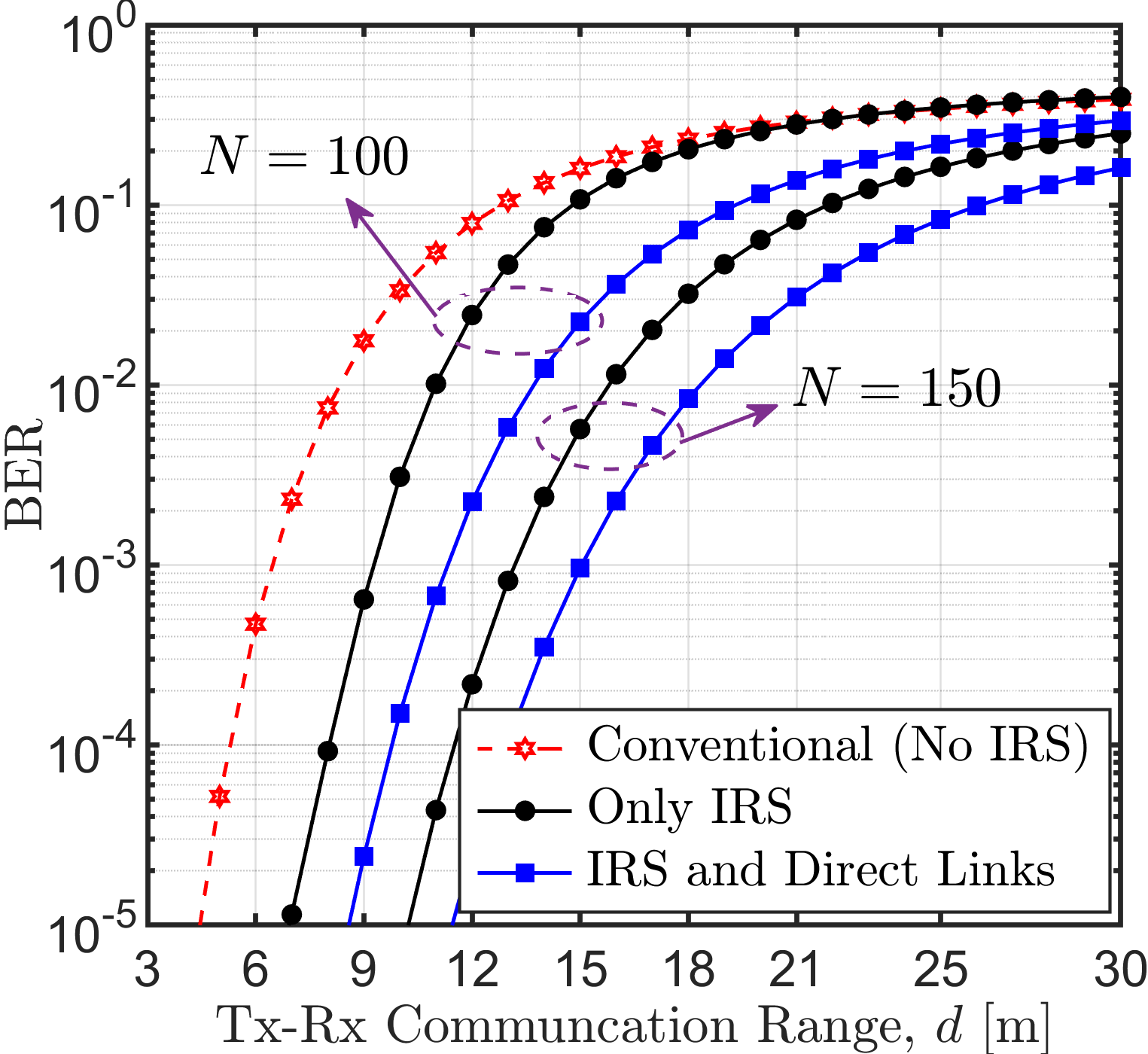}\par \vspace{-10pt}\caption{\ Communication range of conventional and RIS systems. }\label{Fig8}
\end{multicols}\vspace{-40pt}
\end{figure*}

To evaluate how RISs extend the communication range of noise modulation-based systems, Fig. \ref{Fig8} shows the BER versus Tx-Rx distance for three setups: (i) a conventional system with direct Tx-Rx and interferers-Rx links only, (ii) an RIS-assisted system without direct links, and (iii) an RIS-assisted system with direct links. As shown in Fig. \ref{Fig7}, the Rx and RIS are fixed, with the RIS mounted at a height of $H=3$ m. Accordingly, the RIS-Rx, interferers-Rx, and interferers-RIS distances are fixed, as previously described. We assume that the Tx moves horizontally to test the range extension. Using the Pythagorean theorem, the fixed horizontal RIS-Rx distance is $\chi = \sqrt{d_{2}^{2} - H^{2}}$, and the variable Tx-RIS distance is $d_{1} = \sqrt{H^{2} + (d - \chi)^{2}}$. The REs are equally split between EH and signal reflection ($N_{1}=N_{2}=N/2$), with ${\sigma_{0}^2,P_{k}}={5,20}$ dBm, and ${\mathcal{R},K,M}={3,4,15}$. Fig. \ref{Fig8} reveals that for a target BER, integrating an RIS with noise modulation significantly improves the Tx-Rx communication range, {where the performance gain is further} improved by increasing the number of REs. For example, for a BER of $10^{-3}$, the RIS-assisted systems with $N=100$ achieve the Tx-Rx communication ranges of $d\approx 9.5$ m (without direct links) and $d\approx 11.5$ m (with direct links). These greatly outperform the conventional system without an RIS, which has a range of $d\approx 6.5$ m. Increasing the number of REs to $N=150$ further extends the range to $d\approx 13.4$ m for the system without direct links, and to $d=15$ m for the system with direct links. The improved performance results from the additional communication paths created by the RIS, leading to a boost in the received signal power.

Figs. \ref{FigAsyECSR}, \ref{FigAsyBER}, and \ref{FigAsyMI} show the ECSR, BER, and MI versus interference power for various numbers of interferers {and EH models}. Here, REs are equally split for EH and reflection, i.e., $N_{1}=N_{2}=N/2$, with $M=10$, $\sigma_{0}^{2}=-10$ dBm, $\mathcal{R}=5$, and $\{d_{\text{D}}^{(k)},d_{1}^{(k)}\}=\{10,5\}$ m for $k=1,\dots,K$; remaining parameters follow the first paragraph of this section. In the low-interference regime, performance remains nearly unchanged. For instance, {the LEH system} with $K=5$ yields an ECSR of 0.009 at $P_k=5$ dBm, corresponding to a BER of 0.34 and MI of 0.24 nats. This is due to insufficient harvested energy to meet the constraints and enable RIS beamforming, causing REs to reflect signals blindly, with minimal performance gain.

{It is also observed that for the same numbers of REs and interferers, the performance maximization under NLEH requires a higher interference power. For instance, with $K = 5$, the ECSR under NLEH reaches its maximum at $P_k \geq 17$ dBm, whereas under the LEH, it peaks at $P_k \geq 12$ dBm. This discrepancy arises from the nonlinear characteristics of the NLEH circuitry. This variation in energy availability has a direct impact on the system performance, as reflected in the BER and MI in Figs. \ref{FigAsyBER} and \ref{FigAsyMI}. However, beyond EH, interference also acts as a direct impairment to the received signal. Although increasing $P_{k}$ improves the RIS beamforming via enhanced EH, it also degrades the communication performance by raising interference at the Rx. This dual effect leads to distinct optimal points in performance metrics. For example, under LEH, the BER with $K=5$ reaches its minimum of $1.2\times 10^{-4}$ at $P_k \approx 12$ dBm and the MI peaks at $2.56$ nats. In contrast, under NLEH, the minimum BER of $1.2\times 10^{-2}$ and peak MI of $1.33$ nats occur at $P_k \approx 17$ dBm. These shifts are due to the fact that NLEH requires more interference to energize the RIS, which in turn further degrades signal quality at the Rx.} Moreover, the EH performance is maximized by increasing the interference power and the number of interferers. For instance, as shown in Fig. \ref{FigAsyECSR}, the ECSR of the system with LEH and $K=5$ reaches its maximum at $P_{k}\geq 12$ dBm, {while it reaches the same value at $P_{k}\geq 5$ dBm for $K=15$ and $P_{k}\geq 2$ dBm for $K=30$,} validating equation (\ref{asyeh}).
\begin{figure*}[t]
  \centering
  \begin{subfigure}[b]{0.32\textwidth}
    \centering
    \includegraphics[width=\linewidth]{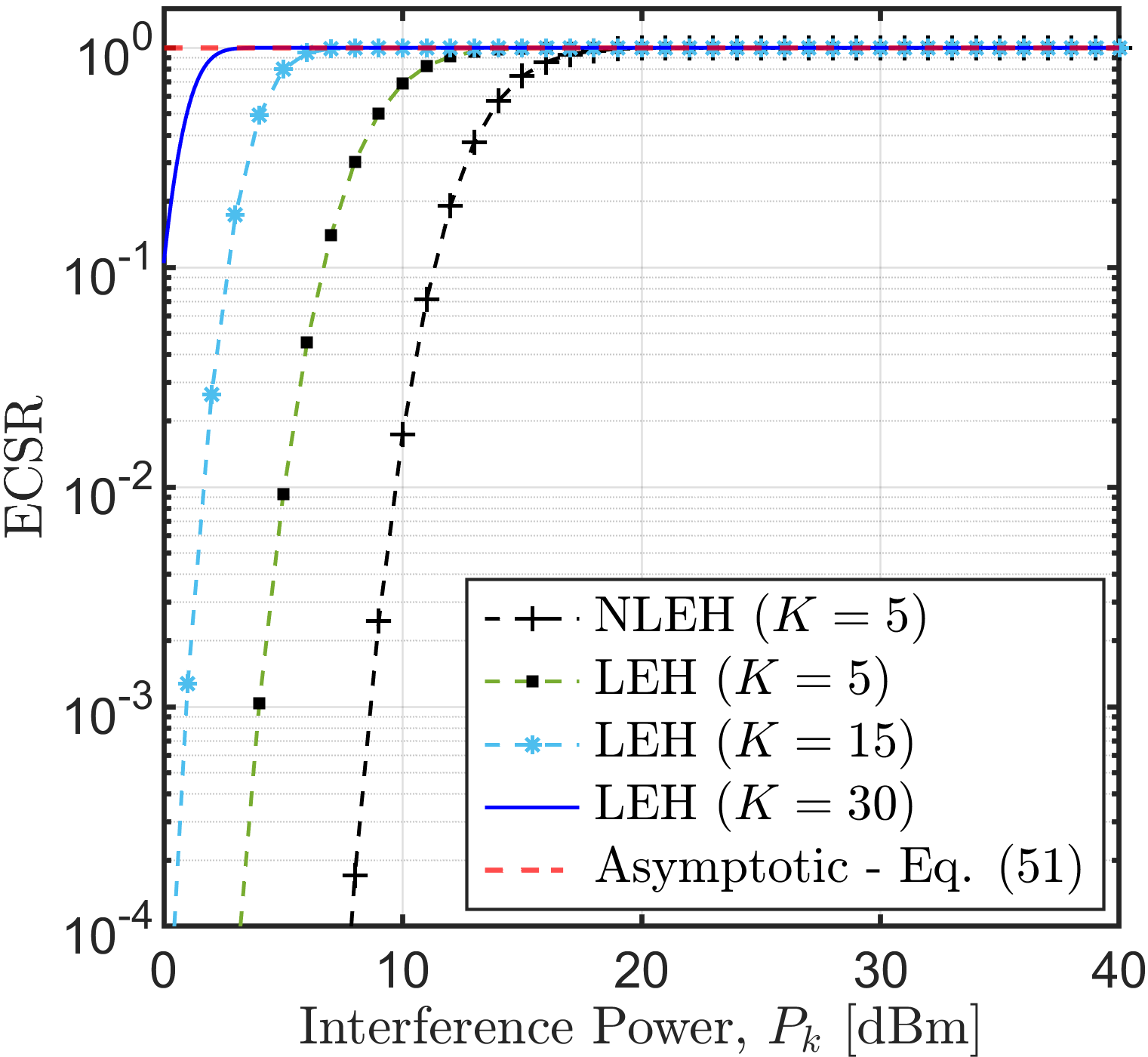}
    \vspace{-25pt}\caption{}
    \label{FigAsyECSR}
  \end{subfigure}
  \hfill
  \begin{subfigure}[b]{0.32\textwidth}
    \centering
    \includegraphics[width=\linewidth]{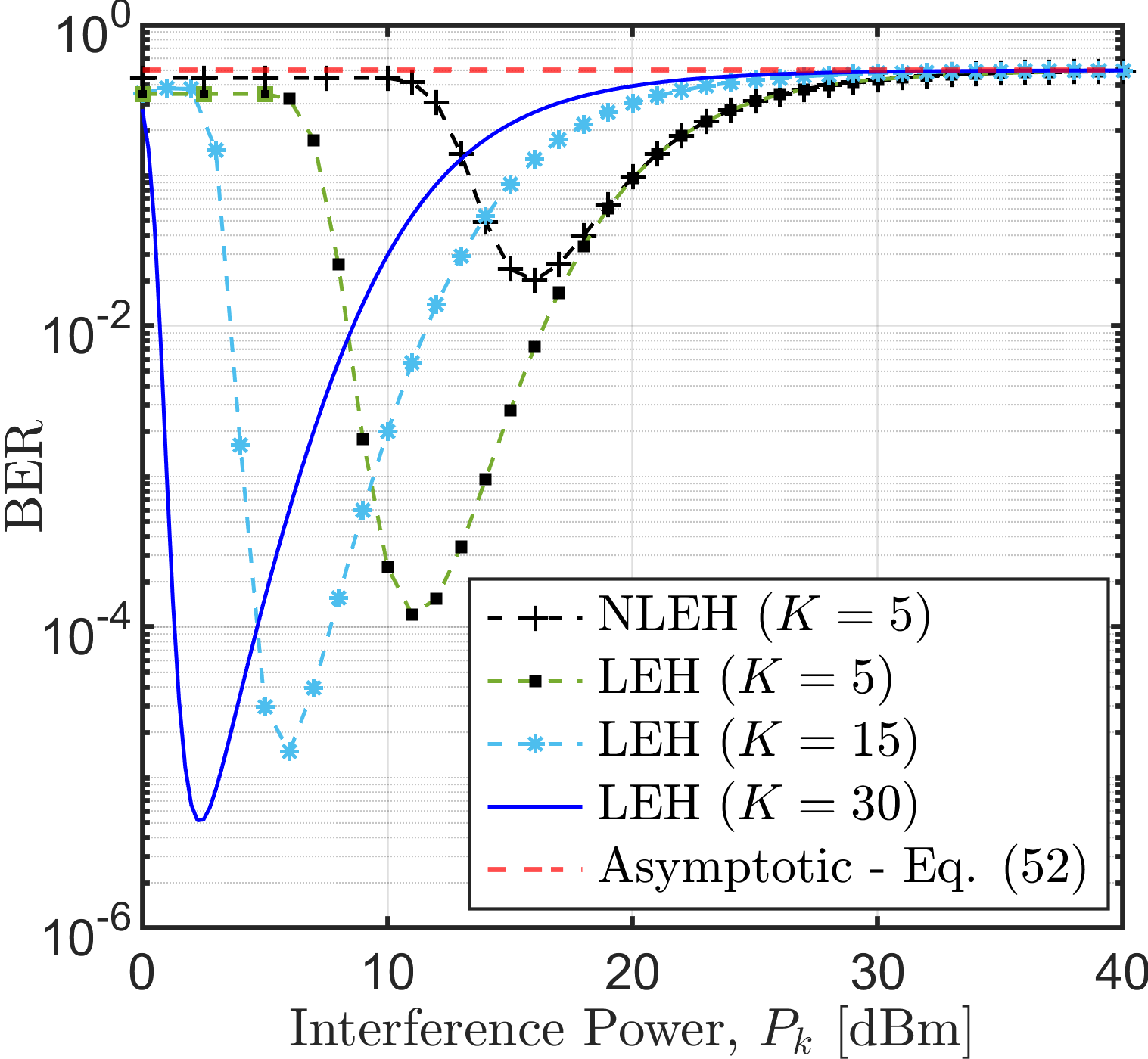}
    \vspace{-25pt}\caption{}
    \label{FigAsyBER}
  \end{subfigure}
  \hfill
  \begin{subfigure}[b]{0.32\textwidth}
    \centering
    \includegraphics[width=\linewidth]{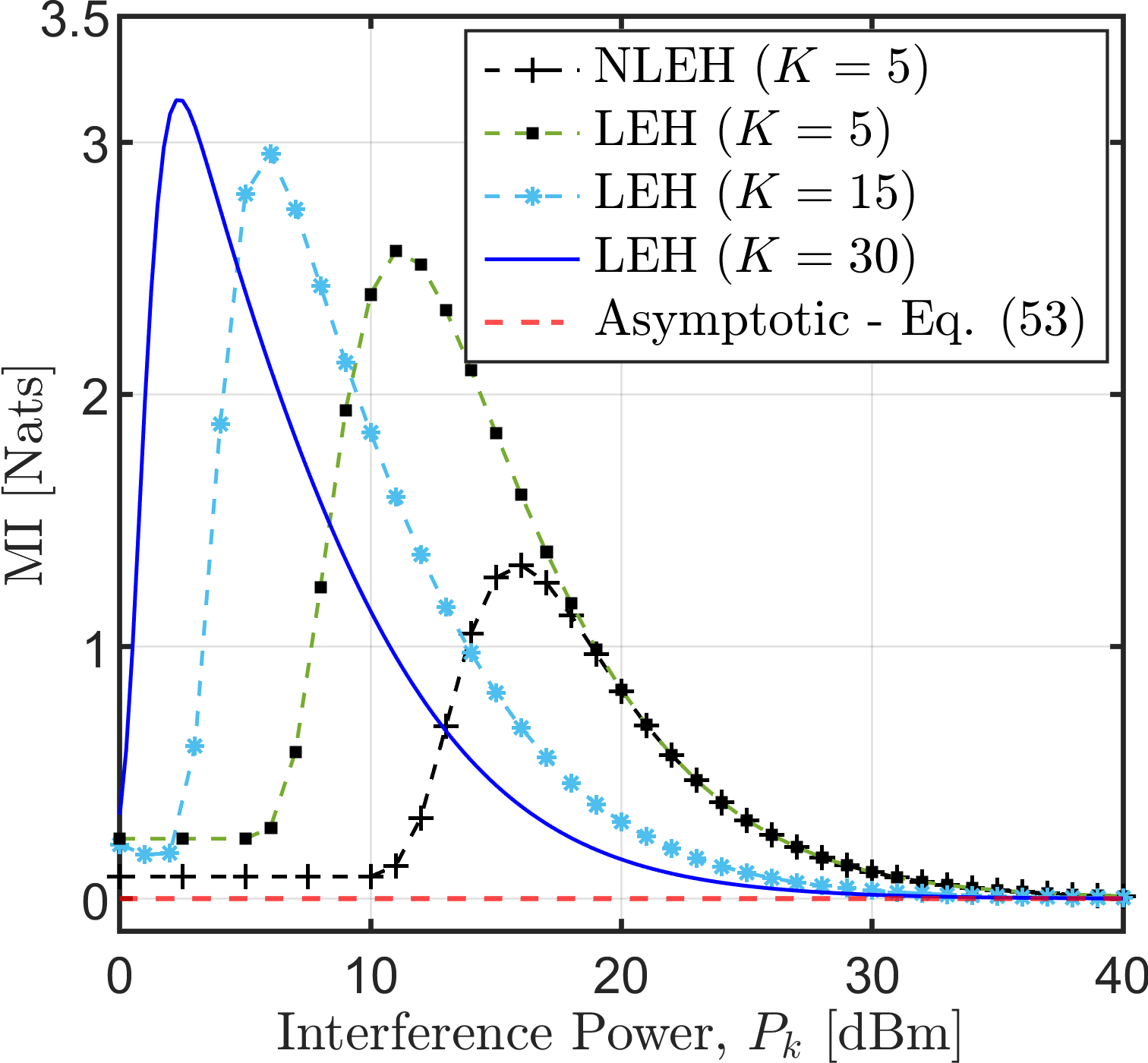}
    \vspace{-25pt}\caption{}
    \label{FigAsyMI}
  \end{subfigure}
\vspace{-15pt}\caption{\ {Impact of different interference levels on the system performance: (a) ECSR, (b) BER, and (c) MI.}}
  \label{FigAllAsymp}\vspace{-30pt}
\end{figure*}

Furthermore, the communication performance improves when moving from the low to a moderate-interference regime, where the ECSR is maximized with the lowest necessary interference power. For instance, under LEH, when $K=5$ and $P_{k}$ is increased to 12 dBm, the ECSR reaches its maximum with minimal interference effects on communication, achieving the lowest BER and highest MI. This occurs because an ECSR of one indicates that sufficient power is available for RIS beamforming, enabling the desired signals to be focused towards the Rx. {However, further increasing $P_{k}$ beyond 12 dBm or $K$ to 30 leads to a deterioration in the BER and MI, rapidly reaching their worst values of 0.5 and 0 nats, respectively, which validates (\ref{ASBER}), (\ref{ASMI}), and Remark \ref{REM3}.} This is because once the energy requirement is met at $P_{k}=12$ dBm {or at a lower power with $K=30$,} any additional energy harvested by the RIS becomes ineffective. Moreover, for $P_{k} > 12$ dBm and a large $K$, the extra interference power reaching the Rx through both the direct and RIS links negatively affects the communication, making the additional energy useless. Although the RIS phase shifts are configured to beamform the desired signals and suppress the interference through random reflections towards the Rx\cite{10328394,9440664,10438079}, the direct interference links to the Rx still exist. Therefore, in high-interference scenarios where $\{K,P_{k}\}\to\infty$, designing efficient interference cancellation techniques at the Rx becomes necessary. 
\vspace{-10pt}\section{Conclusion}\label{SECCon}
\vspace{-10pt}In this paper, we investigated the integration of a {zero-energy} RIS with noise modulation as an energy-efficient solution for IoT communications. By exploiting the co-existing interference signals as a potential energy source for RIS-based EH, the proposed system eliminates the need for dedicated power sources. We presented a closed-form solution and an algorithm to optimize the allocation of REs between EH and signal reflection, considering the practical constraints of phase quantization errors, random energy availability and random beamforming capability. We also derived performance metrics, including the ECSR, BER, optimal threshold, MI, and EE, providing valuable insights into the trade-offs between EH and communication performance.

The results demonstrated that increasing the number of REs allocated for EH improves the ECSR and communication performance in moderate interference regimes. However, excessive RE allocation to EH degrades the communication performance due to reduced signal reflection capability. Moreover, {this paper emphasized the importance of interference levels}, showing that while interference signals can enhance the EH and communication in low and moderate interference regimes, they become detrimental in high-interference scenarios. In conclusion, this work demonstrated the viability of integrating noise modulation with {zero-energy} RISs for enhancing the range, reliability, and EE of IoT systems. It also highlighted that future research is required to focus on advanced RIS designs and interference cancellation techniques to further improve the performance in high-interference environments. {Furthermore, an important direction for future research is the real-time adaptation of the system to dynamic environmental conditions, particularly the effects of time-varying dynamic channels on the system performance.} 
\appendices
\section{}
Let's first express the complex Gaussian channels, $\hat{h}_{\ell}$ and $\hat{g}_{\ell}$, in terms of their real and imaginary components as $\hat{h}_{\ell}=\hat{h}_{\ell}^{\text{R}}+j\hat{h}_{\ell}^{\text{I}}$ and $\hat{g}_{\ell}=\hat{g}_{\ell}^{\text{R}}+j\hat{g}_{\ell}^{\text{I}}$, where $\{\hat{h}_{\ell}^{\text{R}},\hat{h}_{\ell}^{\text{I}},\hat{g}_{\ell}^{\text{R}},\hat{g}_{\ell}^{\text{I}}\}\sim\mathcal{N}\left(0,1/2\right)$. Thus, $\mathcal{D}$ given in (\ref{Dp}) can be written as $\mathcal{D}\triangleq\mathcal{D}_{\text{R}}^{2}+\mathcal{D}_{\text{I}}^{2}$, where $\mathcal{D}_{\text{R}}$ and $\mathcal{D}_{\text{I}}$ are expressed as
\begin{align}
&\mathcal{D}_{\text{R}}=\mathcal{L}_{\text{d}}|h_{\text{d}}|+\mathcal{L}_{B}\textstyle\sum_{i=1}^{N_{B}}|h_{i}||g_{i}|\cos\phi_{i}+\mathcal{L}_{B}\textstyle\sum_{\ell=1}^{N_{b}}\hat{h}_{\ell}^{\text{R}}\hat{g}_{\ell}^{\text{R}}.\\
&\mathcal{D}_{\text{I}}=\mathcal{L}_{B}\textstyle\sum_{i=1}^{N_{B}}|h_{i}||g_{i}|\sin\phi_{i}+\mathcal{L}_{B}\textstyle\sum_{\ell=1}^{N_{b}}\hat{h}_{\ell}^{\text{I}}\hat{g}_{\ell}^{\text{I}}.
\end{align}

Therefore, by exploiting \cite{10328394} and \cite{10047976}, the mean and variance of $\mathcal{D}$ can be expressed as 
\begin{align}
&\mu_{\mathcal{D}}=\mathbb{E}\left[\mathcal{D}_{\text{R}}^{2}\right]+\mathbb{E}\left[\mathcal{D}_{\text{I}}^{2}\right]=\mathbb{V}\left[\mathcal{D}_{\text{R}}\right]+\mathbb{V}\left[\mathcal{D}_{\text{I}}\right]+\mathbb{E}^{2}\left[\mathcal{D}_{\text{R}}\right]+\mathbb{E}^{2}\left[\mathcal{D}_{\text{I}}\right].\\
&\sigma_{\mathcal{D}}^{2}=\mathbb{V}\left[\mathcal{D}_{\text{R}}^{2}\right]+\mathbb{V}\left[\mathcal{D}_{\text{I}}^{2}\right]=2\mathbb{V}\left[\mathcal{D}_{\text{R}}\right](\mathbb{V}\left[\mathcal{D}_{\text{R}}\right]+2\mathbb{E}^{2}\left[\mathcal{D}_{\text{R}}\right])+2\mathbb{V}\left[\mathcal{D}_{\text{I}}\right](\mathbb{V}\left[\mathcal{D}_{\text{I}}\right]+2\mathbb{E}^{2}\left[\mathcal{D}_{\text{I}}\right]).
\end{align}

To find $\mu_{\mathcal{D}}$ and $\sigma_{\mathcal{D}}^{2}$, the mean and variance of $\mathcal{D}_{\text{R}}$ and $\mathcal{D}_{\text{I}}$ are required, which are written as
\begin{align}
&\mathbb{E}\left[\mathcal{D}_{\text{R}}\right]=\mathcal{L}_{\text{d}}\mathbb{E}\left[|h_{\text{d}}|\right]+\mathcal{L}_{B}\mathbb{E}\left[\mathcal{G}_{1}\right]+\mathcal{L}_{B}\mathbb{E}\left[\mathcal{G}_{2}\right],\\
&\mathbb{V}\left[\mathcal{D}_{\text{R}}\right]=\mathcal{L}_{\text{d}}^{2}\mathbb{V}\left[|h_{\text{d}}|\right]+\mathcal{L}_{B}^{2}\mathbb{V}\left[\mathcal{G}_{1}\right]+\mathcal{L}_{B}^{2}\mathbb{V}\left[\mathcal{G}_{2}\right],
\end{align}
\begin{align}
    &\mathbb{E}\left[\mathcal{D}_{\text{I}}\right]=\mathcal{L}_{B}\mathbb{E}\left[\mathcal{G}_{3}\right]+\mathcal{L}_{B}\mathbb{E}\left[\mathcal{G}_{4}\right],\\
&\mathbb{V}\left[\mathcal{D}_{\text{I}}\right]=\mathcal{L}_{B}^{2}\mathbb{V}\left[\mathcal{G}_{3}\right]+\mathcal{L}_{B}^{2}\mathbb{V}\left[\mathcal{G}_{4}\right],
\end{align}
where $\mathcal{G}_{1}\triangleq\sum_{i=1}^{N_{B}}A_{i}$ with $A_{i}=|h_{i}||g_{i}|\cos\phi_{i}$, $\mathcal{G}_{2}\triangleq\sum_{\ell=1}^{N_{b}}\hat{h}_{\ell}^{\text{R}}\hat{g}_{\ell}^{\text{R}}$, $\mathcal{G}_{3}\triangleq\sum_{i=1}^{N_{B}}B_{i}$ with $B_{i}=|h_{i}||g_{i}|\sin\phi_{i}$, and $\mathcal{G}_{4}\triangleq\sum_{\ell=1}^{N_{b}}\hat{h}_{\ell}^{\text{I}}\hat{g}_{\ell}^{\text{I}}$.

Since $h_{\text{d}}\sim\mathcal{CN}\left(\mu_{\text{d}},1\right)$ is a non-zero mean complex Gaussian RV, $|h_{\text{d}}|$ is a Rician RV with $\mathbb{E}\left[|h_{\text{D}}|\right]=\frac{\sqrt{\pi}}{2}L_{\frac{1}{2}}\left(-\mu_{\text{d}}^{2}\right)$ and $\mathbb{V}\left[|h_{\text{D}}|\right]=L_{1}\left(-\mu_{\text{d}}^{2}\right)-\frac{\pi}{4}L_{\frac{1}{2}}^{2}\left(-\mu_{\text{d}}^{2}\right)$ \cite{10047976}. Moreover, as the numbers of REs capable of beamforming and blindly reflecting the signals, $N_{B}$ and $N_{b}$, are binomial RVs with $\mathbb{E}[N_{B}]=N_{2}{\mathcal{P}_{s}^{\zeta}}$, $\mathbb{E}[N_{b}]=N_{2}(1-{\mathcal{P}_{s}^{\zeta}})$, and $\mathbb{V}[N_{B}]=\mathbb{V}[N_{b}]=N_{2}{\mathcal{P}_{s}^{\zeta}}(1-{\mathcal{P}_{s}^{\zeta}})$, the RVs $\{\mathcal{G}_{1},\mathcal{G}_{2},\mathcal{G}_{3},\mathcal{G}_{4}\}$ are compound RVs\cite{ROSS}. Thus, using \cite[Ch. 3]{ROSS}, we can write that
\begin{align}
    &\mathbb{E}\left[\mathcal{G}_{1}\right]=\mathbb{E}\left[\textstyle\sum_{i=1}^{N_{B}}A_{i}\right]=\mathbb{E}[N_{B}]\mathbb{E}[A_{i}]=N_{2}{\mathcal{P}_{s}^{\zeta}}\mathbb{E}[A_{i}],\\
 & \mathbb{V}\left[\mathcal{G}_{1}\right]=\mathbb{V}\left[\textstyle\sum_{i=1}^{N_{B}}A_{i}\right]=\mathbb{E}[N_{B}]\mathbb{V}[A_{i}]+\mathbb{V}[N_{B}]\mathbb{E}^{2}[A_{i}]=N_{2}{\mathcal{P}_{s}^{\zeta}}\mathbb{V}[A_{i}]+N_{2}{\mathcal{P}_{s}^{\zeta}}(1-{\mathcal{P}_{s}^{\zeta}})\mathbb{E}^{2}[A_{i}],
\end{align}
\vspace{-10pt}\begin{align}
   & \mathbb{E}\left[\mathcal{G}_{2}\right]=\mathbb{E}\left[\textstyle\sum_{\ell=1}^{N_{b}}\hat{h}_{\ell}^{\text{R}}\hat{g}_{\ell}^{\text{R}}\right]=\mathbb{E}[N_{b}]\mathbb{E}\left[\hat{h}_{\ell}^{\text{R}}\hat{g}_{\ell}^{\text{R}}\right]=N_{2}(1-{\mathcal{P}_{s}^{\zeta}})\mathbb{E}\left[\hat{h}_{\ell}^{\text{R}}\hat{g}_{\ell}^{\text{R}}\right]=0,\\
    &\mathbb{V}\left[\mathcal{G}_{2}\right]=\mathbb{V}\left[\textstyle\sum_{\ell=1}^{N_{b}}\hat{h}_{\ell}^{\text{R}}\hat{g}_{\ell}^{\text{R}}\right]=\mathbb{E}[N_{b}]\mathbb{V}\left[\hat{h}_{\ell}^{\text{R}}\hat{g}_{\ell}^{\text{R}}\right]+\mathbb{V}[N_{b}]\mathbb{E}^{2}\left[\hat{h}_{\ell}^{\text{R}}\hat{g}_{\ell}^{\text{R}}\right]=\frac{N_{2}(1-{\mathcal{P}_{s}^{\zeta}})}{4},
\end{align}
\begin{align}
    &\mathbb{E}\left[\mathcal{G}_{3}\right]=\mathbb{E}\left[\textstyle\sum_{i=1}^{N_{B}}B_{i}\right]=N_{2}{\mathcal{P}_{s}^{\zeta}}\mathbb{E}[B_{i}],\\
    &\mathbb{V}\left[\mathcal{G}_{3}\right]=\mathbb{V}\left[\textstyle\sum_{i=1}^{N_{B}}B_{i}\right]=N_{2}{\mathcal{P}_{s}^{\zeta}}\mathbb{V}[B_{i}]+N_{2}{\mathcal{P}_{s}^{\zeta}}(1-{\mathcal{P}_{s}^{\zeta}})\mathbb{E}^{2}[B_{i}],\\
   & \mathbb{E}\left[\mathcal{G}_{4}\right]=\mathbb{E}\left[\textstyle\sum_{\ell=1}^{N_{b}}\hat{h}_{\ell}^{\text{I}}\hat{g}_{\ell}^{\text{I}}\right]=N_{2}(1-{\mathcal{P}_{s}^{\zeta}})\mathbb{E}\left[\hat{h}_{\ell}^{\text{I}}\hat{g}_{\ell}^{\text{I}}\right]=0,\\   &\mathbb{V}\left[\mathcal{G}_{4}\right]=\mathbb{V}\left[\textstyle\sum_{\ell=1}^{N_{b}}\hat{h}_{\ell}^{\text{I}}\hat{g}_{\ell}^{\text{I}}\right]=\mathbb{E}[N_{b}]\mathbb{V}\left[\hat{h}_{\ell}^{\text{I}}\hat{g}_{\ell}^{\text{I}}\right]+\mathbb{V}[N_{b}]\mathbb{E}^{2}\left[\hat{h}_{\ell}^{\text{I}}\hat{g}_{\ell}^{\text{I}}\right]=\frac{N_{2}(1-{\mathcal{P}_{s}^{\zeta}})}{4},
\end{align}
where $\mathbb{E}\left[\hat{h}_{\ell}^{\text{R}}\hat{g}_{\ell}^{\text{R}}\right]=\mathbb{E}\left[\hat{h}_{\ell}^{\text{I}}\hat{g}_{\ell}^{\text{I}}\right]=0$, $\mathbb{V}\left[\hat{h}_{\ell}^{\text{R}}\hat{g}_{\ell}^{\text{R}}\right]=\mathbb{V}\left[\hat{h}_{\ell}^{\text{I}}\hat{g}_{\ell}^{\text{I}}\right]=1/4$, and the mean and variance of $A_{i}$ and $B_{i}$ are readily available in different studies of RISs, which are written as $\mathbb{E}\left[A_{i}\right]=Q\sin\left(\pi/Q\right)/4$, $\mathbb{E}\left[B_{i}\right]=0$, $\mathbb{V}\left[A_{i}\right]=1/2+Q\sin\left(2\pi/Q\right)/(4\pi)-Q^{2}\sin^{2}\left(\pi/Q\right)/16$ and $\mathbb{V}\left[B_{i}\right]=1/2-Q\sin\left(2\pi/Q\right)/(4\pi)$\cite{10328394,10047976}. 

Finally, by substituting the corresponding mean and variance terms, and after some algebraic manipulations, $\mu_{\mathcal{D}}$ and $\sigma_{\mathcal{D}}^{2}$ given in (\ref{MEAND}) and (\ref{VARD}) are obtained. $\hfill\blacksquare$ 
\balance
\vspace{-10pt}\bibliographystyle{IEEEtran} 
\bibliography{abr,ref}

\end{document}